\documentclass[twocolumn]{aastex63}
\usepackage{amsmath}

\newcommand{\msun}{${\rm M_{\sun}}$}

\def\ltsima{$\; \buildrel < \over \sim \;$}
\def\simlt{\lower.5ex\hbox{\ltsima}}
\def\gtsima{$\; \buildrel > \over \sim \;$}
\def\simgt{\lower.5ex\hbox{\gtsima}}
\def\kms{{\rm\,km\,s^{-1}}}
\def\mas{{\rm\,mas}}
\def\masyr{{\rm\,mas/yr}}
\def\kpc{{\rm\,kpc}}

\def\msun{{\rm\,M_\odot}}

\def\pc{{\rm\,pc}}

\def\g{{\rm\,g}}

\def \r {r$^{1/4}$ }


\def\deg{^\circ}
\def\degg{\hbox{$\null^\circ$\hskip-3pt .}}

\def\s{\ifmmode \widetilde \else \~\fi}
\def\={\overline}

\def\spose#1{\hbox to 0pt{#1\hss}}

\def\lta{\mathrel{\spose{\lower 3pt\hbox{$\mathchar"218$}}
     \raise 2.0pt\hbox{$\mathchar"13C$}}}
\def\gta{\mathrel{\spose{\lower 3pt\hbox{$\mathchar"218$}}
     \raise 2.0pt\hbox{$\mathchar"13E$}}}
\def\Dt{\spose{\raise 1.5ex\hbox{\hskip3pt$\mathchar"201$}}}    
\def\dt{\spose{\raise 1.0ex\hbox{\hskip2pt$\mathchar"201$}}}    

\def\r{${\rm r^{1/4}}$}

\def\dotsfill{\leaders\hbox to 1em{\hss.\hss}\hfill}
\def\sun{\odot}
\def\earth{\oplus}

\def\Gyr{{\rm\,Gyr}}

\def\ltsima{$\; \buildrel < \over \sim \;$}
\def\gtsima{$\; \buildrel > \over \sim \;$}
\def\lsim{\lower.5ex\hbox{\ltsima}}
\def\gsim{\lower.5ex\hbox{\gtsima}}
\def\lapp{\ifmmode\stackrel{<}{_{\sim}}\else$\stackrel{<}{_{\sim}}$\fi}
\def\gapp{\ifmmode\stackrel{>}{_{\sim}}\else$\stackrel{<}{_{\sim}}$\fi}

\def\yC{y_{\rm C}}
\def\xE{x_{\rm E}}
\def\yobs{y_{\rm obs}}

\accepted{February 3, 2020}

\shorttitle{Gaps and Density Spikes in the GD-1 Stream}
\shortauthors{Ibata et al.}
\graphicspath{{./}{figures/}}

\begin{document}

\title{Detection of Strong Epicyclic Density Spikes in the GD-1 Stellar Stream \\
An Absence of Evidence for the Influence of Dark Matter Subhalos?}

\correspondingauthor{Rodrigo Ibata}
\email{rodrigo.ibata@astro.unistra.fr}

\author[0000-0002-3292-9709]{Rodrigo Ibata}
\affiliation{Universit\'e de Strasbourg, CNRS, Observatoire astronomique de Strasbourg, UMR 7550, F-67000 Strasbourg, France}
\nocollaboration{1}

\author[0000-0002-2468-5521]{Guillaume Thomas}
\affiliation{NRC Herzberg Astronomy and Astrophysics, 5071 West Saanich Road, Victoria, BC V9E 2E7, Canada}
\nocollaboration{1}

\author[0000-0003-3180-9825]{Benoit Famaey}
\affiliation{Universit\'e de Strasbourg, CNRS, Observatoire astronomique de Strasbourg, UMR 7550, F-67000 Strasbourg, France}
\nocollaboration{1}

\author[0000-0002-8318-433X]{Khyati Malhan}
\affiliation{The Oskar Klein Centre, Department of Physics, Stockholm University, AlbaNova, SE-10691 Stockholm, Sweden}
\nocollaboration{1}

\author[0000-0002-1349-202X]{Nicolas Martin}
\affiliation{Universit\'e de Strasbourg, CNRS, Observatoire astronomique de Strasbourg, UMR 7550, F-67000 Strasbourg, France}
\affiliation{Max-Planck-Institut f\"ur Astronomie, K\"onigstuhl 17, D-69117, Heidelberg, Germany}
\nocollaboration{1}

\author[0000-0002-6863-0661]{Giacomo Monari}
\affiliation{Universit\'e de Strasbourg, CNRS, Observatoire astronomique de Strasbourg, UMR 7550, F-67000 Strasbourg, France}
\nocollaboration{1}

\begin{abstract}
The density variations in thin stellar streams may encode important information on the nature of the dark matter. For instance, if dark matter aggregates into massive sub-halos, these perturbers are expected to scatter stars out of dynamically cold stellar streams, possibly leading to detectable gaps in those structures. Here we re-examine the density variations in the GD-1 stream, using Gaia DR2 astrometry, Pan-STARRS photometry, together with high precision radial velocities measured with the CFHT/ESPaDOnS and VLT/UVES instruments and complemented with public radial velocity catalogs. We show that after correcting for projection effects, the density profile exhibits high contrast periodic peaks, separated by $2.64\pm0.18\kpc$. An N-body simulation is presented that reproduces this striking morphology with simple epicyclic motion in a smooth Galactic potential. We also discuss the reliability of measuring density variations using ground-based photometric surveys, and for the particular case of GD-1 we highlight some of the artifacts present in the Gaia DR2 catalog along its track. Massive dark subhalos do not appear to be required to explain the density clumping along GD-1.
\end{abstract}


\section{Introduction}
\label{sec:Introduction}

A key prediction of $\Lambda$ Cold Dark Matter ($\Lambda$CDM) cosmology is that galaxies reside within dark matter halos that are composed of a hierarchy of smaller ``sub-halos''. This hierarchy is expected to continue down in mass all the way to a limit set by the (as yet unknown) thermal free-streaming length of the dark matter particle \citep{2008MNRAS.391.1685S}. Many thousands of these sub-halos are expected to orbit within our Galaxy, but only the most massive would contain some baryonic component that could render them directly observable (as satellite galaxies). So detecting the huge predicted population of completely dark sub-halos requires identifying their gravitational influence on photons or on observable baryonic structures. 

One promising avenue to detect the sub-halo population is to analyse the morphology and flux-ratios of strongly-lensed quasar images. At present the evidence appears consistent with $\Lambda$CDM \citep{2019MNRAS.485.2179R,2019MNRAS.tmp.2780H}, but at low statistical significance.

It is interesting therefore to consider how the sub-halos may influence stellar sub-structures of our Milky Way or of other nearby galaxies. Although the expected fully-dark sub-halos could be very massive (up to $\sim 10^8\msun$), their large physical scale makes these bodies very ``fluffy'', and interactions with the baryonic components of a galaxy will be subtle. One therefore needs to identify some dynamical probes that respond in a measurable way to small perturbations of the acceleration field. This realisation led several groups to propose that the fragility of dynamically-cold star streams could be used as a means to explore the sub-clustering of the dark matter on sub-galactic scales \citep{2002MNRAS.332..915I,2002ApJ...570..656J,2002MNRAS.336..119M}. 

Heating from a sub-halo fly-by will increase the velocity dispersion in a stream, and given that these initially can be very cold (e.g., the one-dimensional velocity dispersion in the GD-1 stream is $\sim 1\kms$, \citealt{2019MNRAS.486.2995M}), the influence of the sub-halo flyby may be detectable, in principle. However, the practical difficulty in realising such a measurement is that streams generally possess a very low density of stars that are bright enough to be measured with good precision, which makes the dynamical heating effect challenging to detect. 

A promising alternative to measuring velocity dispersion variations (which would require obtaining high-precision line-of-sight kinematics to hundreds or thousands of stars in a stream) is instead to make use of the stream's spatial morphology. \citet{2012ApJ...748...20C} showed that characteristic underdensities or ``gaps'' are formed after a close flyby of a massive perturber. Indeed, for the specific case of the GD-1 stream, \citet{2016ApJ...820...45C} proposed that sub-halos could be responsible for the gaps on scales of $\sim 10\deg$ that were detected in Sloan Digital Sky Survey (SDSS) maps of the system \citep{2013ApJ...768..171C}.

Recently, \citet{2019arXiv191105745D} have remeasured the morphology of the GD-1 stream using the excellent astrometric data from the Second Data Release (DR2) of the Gaia mission \citep{2018A&A...616A...2L,2018A&A...616A...1G} combined with Pan-STARRS photometry \citep{2016arXiv161205560C}. \citet{2019arXiv191102662B} use these data to detect a power spectrum of density variations along the stream that they claim requires the presence of a population of perturbing sub-halos of mass $10^7\msun$ to $10^9\msun$ with a density that is within the uncertainties of $\Lambda$CDM predictions. 

The present work aims to examine these very interesting claims, providing additional data and analysis of the GD-1 system. The layout of this paper is as follows. Section~\ref{sec:GD-1_Stream} presents an overview of the GD-1 system, whose properties we re-derive in Section~\ref{sec:STREAMFINDER_Sample}, based on a clean sample of stars, including new radial velocity measurements. With these new constraints we present the density profile along the stream in Section~\ref{sec:Density_Profile}, finding that the profile is substantially more peaked than found by \citet{2019arXiv191102662B}, and displays periodic overdensities. In Section~\ref{sec:CFHT_Data} we change tack to attempt to quantify the reliability of ground-based photometric surveys in order to estimate how confidently the surface density of a highly contaminated structure can be measured. In Section~\ref{sec:Modelling_the_Density_Profile} we present some simple models to interpret the observed density profile. Our simulations show that the density spikes can be modelled by the escape of stars at low velocity from a globular cluster that has now completely dissolved. Finally, with all these caveats in mind, we measure the power spectrum of the GD-1 stream in Section~\ref{sec:Power_Spectrum_Analysis}. Our conclusions are laid out in Section~\ref{sec:Conclusions}.

\begin{figure*}
\begin{center}
\includegraphics[angle=0, viewport= 25 42 670 640, clip, width=\hsize]{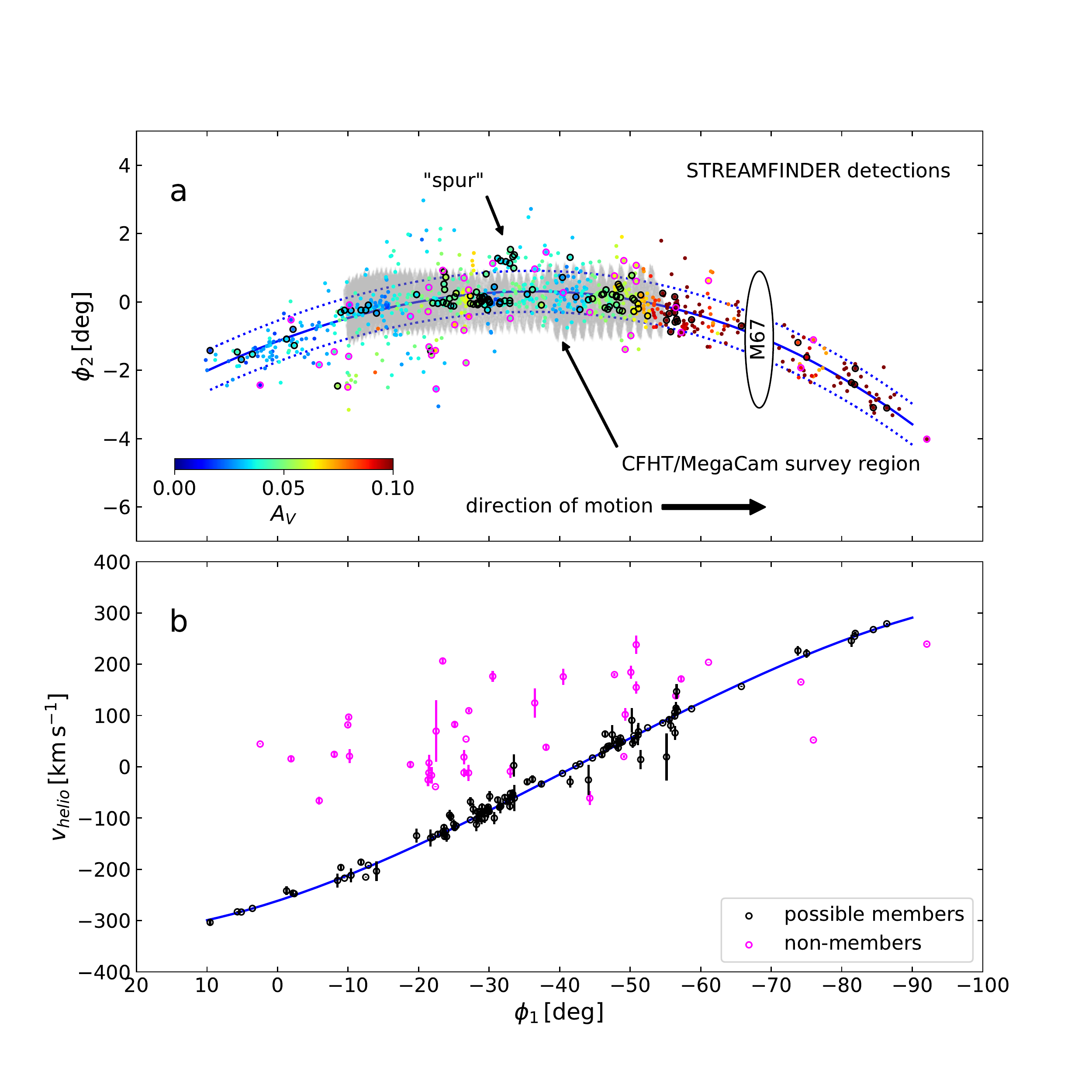}
\end{center}
\caption{a: Spatial distribution of GD-1 stream star detections, using the {\tt STREAMFINDER} software. The $\phi_1,\phi_2$ coordinate system of \citet{2010ApJ...712..260K} is used, where $\phi_1$ points approximately along the stream, while $\phi_2$ is perpendicular to it. The points are color-coded according to the $A_V$ extinction, which can be seen to change in a complex manner along this long structure, possessing wave-like variations on scales of degrees. Additionally, we mark (in gray) the CFHT/MegaCam survey region discussed in Section~\ref{sec:CFHT_Data}. The large foreground open cluster Messier 67 strongly contaminates the GD-1 stream in the marked circular region (which appears elliptical due to the stretching of the $\phi_2$ axis). The GD-1 ``spur'' structure \citep{2018ApJ...863L..20P} is also marked. b: Stars with radial velocities measurements. The magenta open circles show the GD-1 candidates identified by the {\tt STREAMFINDER}, whose measured velocities are incompatible with being GD-1 members. The probable GD-1 stars are displayed in black. These larger open circles (with the same color-coding) are also shown in panel (a). Note that the velocity outliers have a strong tendency to also be spatial outliers. Furthermore, the probable members (based on their radial velocities) define a narrow spatial sequence, with the exception of the ``spur'' grouping. (The error bars show $1\sigma$ velocity uncertainties).} 
\label{fig:map}
\end{figure*}

\section{The GD-1 Stream}
\label{sec:GD-1_Stream}

The GD-1 stream was discovered by \citet{2006ApJ...643L..17G} in the SDSS, where it appeared as a $63\deg$ long structure in matched filter maps designed to reveal metal-poor populations similar to that of the globular cluster M13 (${\rm [Fe/H]}=-1.53$, \citealt{2010arXiv1012.3224H}). The stream lies in the North Galactic cap region in the direction away from the Galactic center. Follow-up medium-resolution spectroscopy obtained by \citet{2010ApJ...712..260K} showed that GD-1 has a relatively circular but retrograde orbit, with a pericenter at $14\kpc$ and an apocenter at $26\kpc$. This orbit keeps the system well away from the inner regions of the Galactic disk, where interactions with giant molecular clouds could cause additional heating \citep{2016MNRAS.463L..17A}, that could contaminate the sought-for signal from $\Lambda$CDM substructure. 

The advent of the Gaia DR2 catalog enabled the search for streams over the full sky using astrometric information in addition to photometry. GD-1 was immediately detected \citep{2018MNRAS.481.3442M} as one of the highest contrast stellar streams in the Galactic halo. Additional stars surrounding the stream were detected \citep{2018ApJ...863L..20P} including an off-track ``spur'' (marked in Figure~\ref{fig:map}a), these features may be revealing the effect of massive perturbers \citep{2019ApJ...880...38B} or they may point to the possibility that the progenitor of GD-1 originated within a larger system \citep{2019ApJ...881..106M}.

\begin{figure}
\begin{center}
\includegraphics[angle=0, viewport= 10 10 400 390, clip, width=\hsize]{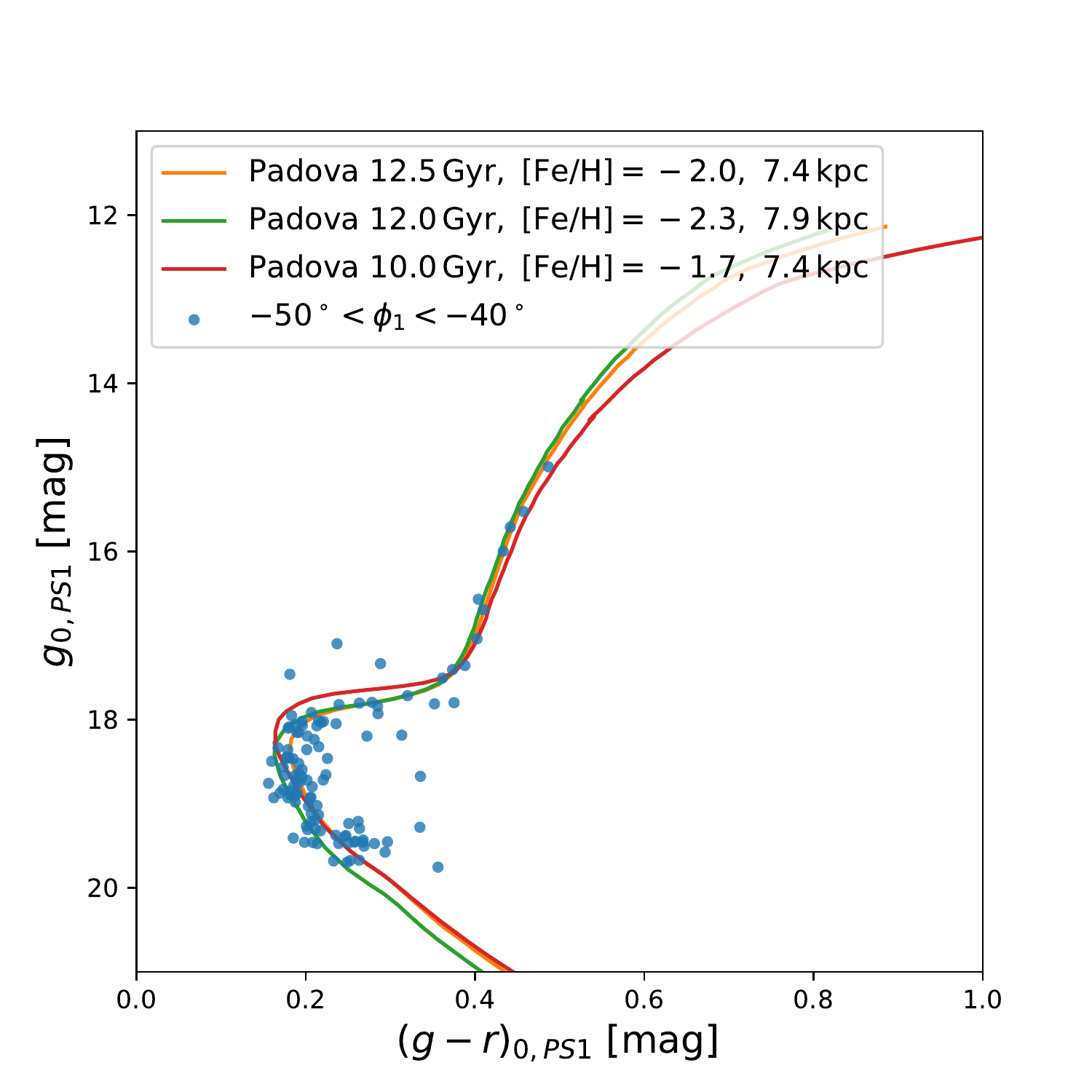}
\end{center}
\caption{Color-magnitude distribution of {\tt STREAMFINDER} stars in the spatial interval $-50\deg<\phi_1<-40\deg$. Similarly well-defined color-magnitude behavior is seen at most locations along the stream, but due to variation of the line of sight distance along the stream, the color-magnitude coherence becomes degraded as data over larger ranges in $\phi_1$ are combined. The selected PARSEC isochrone model with age of $12.5\Gyr$ and metallicity ${\rm [Fe/H]=-2.0}$ provides a plausible representation of these data.}
\label{fig:CMD_and_model}
\end{figure}

\section{The {\tt STREAMFINDER} GD-1 Sample}
\label{sec:STREAMFINDER_Sample}

We aim to derive a sample of GD-1 stars from which we will be able to examine its stellar number density profile. Constructing such a sample is not entirely straightforward though, because of the substantial contamination from normal Galactic field populations over the large area of sky that this structure covers. To extract a clean sample, we first need to know the large-scale behavior of GD-1 in position, parallax, proper motion and photometry.

We will therefore begin our analysis by first deriving the properties of GD-1 from a sample of 811 candidate member stars identified with the {\tt STREAMFINDER} algorithm \citep{2018MNRAS.477.4063M,2018MNRAS.481.3442M}. This software provides a means to assign a likelihood to every star in a dataset according to the possibility of whether the star can be grouped with other stars into a stream-like structure. The adopted algorithm parameters are stated in \citet{2019ApJ...872..152I}; in particular, we searched for stream stars down to $G_0=19.5$~mag using a stream template of Gaussian width $0.05\kpc$, and of length $20\deg$. Three different stellar populations models from the PARSEC library \citep{2012MNRAS.427..127B} were used, with age and metallicity: ($12.5\Gyr, -2.0$), ($12.0\Gyr, -2.3$), and ($10.0\Gyr, -1.7$). For every star we adopted the most likely stream solution obtained from one of these three age-metallicity choices. The resulting spatial distribution of the candidate GD-1 members is shown in Figure~\ref{fig:map}a, displayed in the $\phi_1,\phi_2$ coordinate system of \citet{2010ApJ...712..260K}, where $\phi_1$ corresponds to position on a great circle that is approximately parallel to the GD-1 stream. (For easier comparison to maps in Equatorial coordinates, the $\phi_1$ axis in all figures is displayed such that $\phi_1$ increases towards the left). 

In Figure~\ref{fig:CMD_and_model} we show the color-magnitude diagram (CMD) of a sub-sample of the {\tt STREAMFINDER} detections, using photometry extracted from the second data release (DR2) of the Pan-STARRS survey \citep{2016arXiv161205560C}. Since the stream displays a substantial distance gradient, we selected the sub-sample to lie between $-50\deg<\phi_1<-40\deg$, where the distance is approximately constant. We have chosen to show Pan-STARRS (instead of Gaia) photometry here because of the much smaller uncertainties at the faint end of the CMD. The two PARSEC stellar population models with age and metallicity ($12.5\Gyr, -2.0$) and ($12.0\Gyr, -2.3$) can be seen to give a reasonable representation of GD-1, and are consistent with spectroscopic measurements derived from Segue and LAMOST \citep{2019MNRAS.486.2995M}. We include the more metal-rich model with ($10.0\Gyr, -1.7$) to represent an extreme upper limit to the CMD properties of GD-1.

\begin{figure*}
\begin{center}
\includegraphics[angle=0, viewport= 25 42 670 640, clip, width=\hsize]{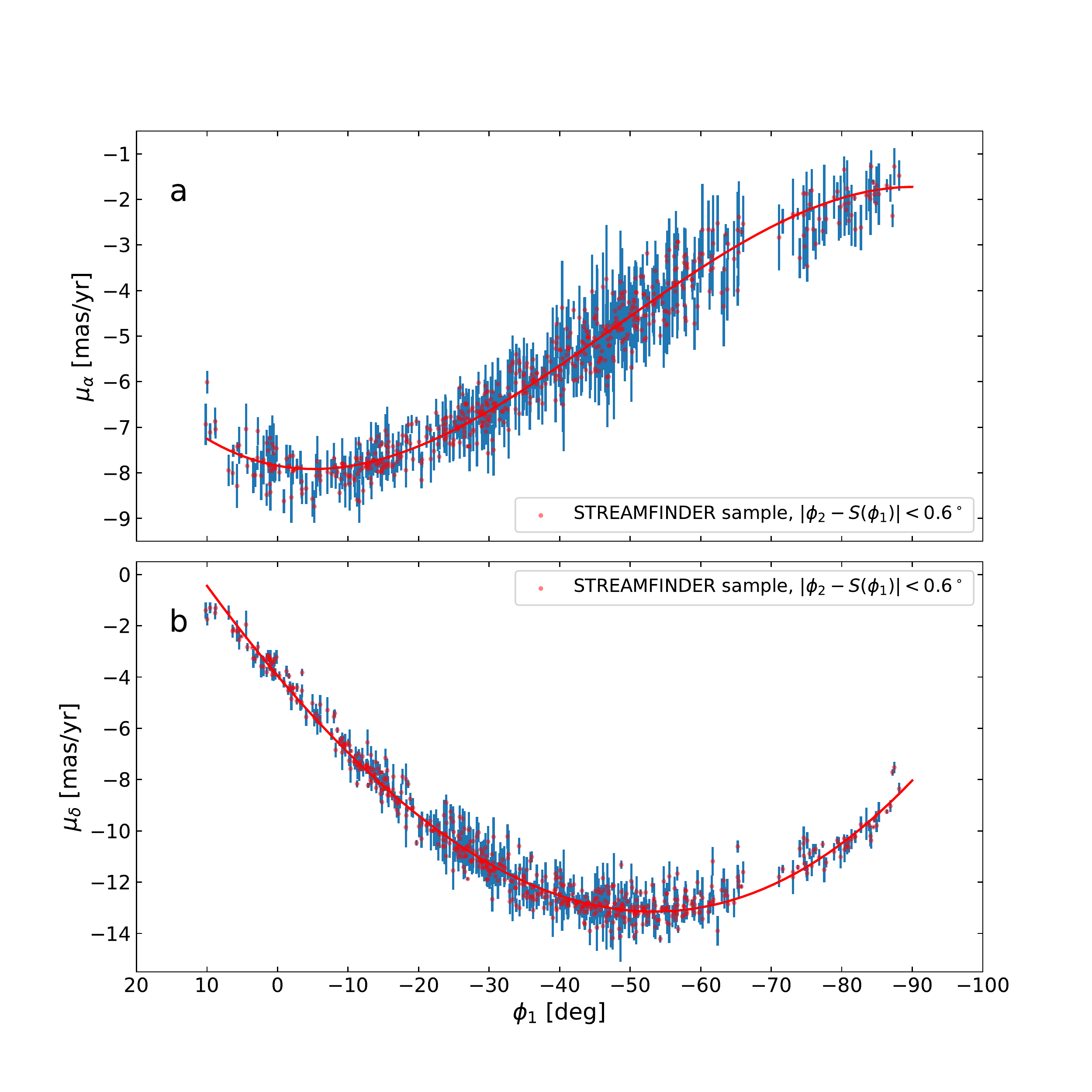}
\end{center}
\caption{Proper motion profiles in $\mu_\alpha$ (a) and $\mu_\delta$ (b). The {\tt STREAMFINDER} sample shown here (containing 603 stars) has been trimmed spatially to lie between the dotted lines in Figure~\ref{fig:map}a, where contamination is low. The fitted cubic polynomials are defined in the text. ($1\sigma$ uncertainties are displayed).}
\label{fig:PMs}
\end{figure*}

As part of an on-going follow-up survey of the stream stars detected with the {\tt STREAMFINDER} algorithm, we observed 29 GD-1 candidate stars with the high-resolution ESPaDOnS spectrograph at the CFHT, 5 stars with the EFOSC2 spectrograph at the NTT, as well as 2 stars with the high-resolution UVES spectrograph at the VLT. The ESPaDOnS spectra were extracted and wavelength-calibrated with the Libre-esprit software \citep{1997MNRAS.291..658D}, and we used the ESOREX pipeline to perform the same task with the EFOSC2 and UVES spectra. The radial velocities of all 35 stars were measured by cross correlation against the radial-velocity standard HD182572 using the ``fxcor'' command in IRAF. The average uncertainty of the stars observed with ESPaDOnS and UVES is $0.9\kms$. An article presenting the spectroscopic follow-up survey of the {\tt STREAMFINDER} detections is currently in preparation, and we defer a detailed exposition of the data to that contribution. 

We cross-matched the {\tt STREAMFINDER} sample against public spectroscopic surveys, finding matches with 2 stars in APOGEE-2 \citep{2017AJ....154...94M}, 2 stars in the Gaia Radial Velocity Spectrometer (RVS) sample, 43 stars in LAMOST DR5 \citep{2012RAA....12.1197C} and 91 stars in SDSS/Segue \citep{2009AJ....137.4377Y}. The final velocity sample (including our CFHT/ESPaDOnS and ESO observations) consists of a total of 156 distinct stars out of the sample of 811. For those stars with multiple measurements, we adopted the measurement that possesses the lowest uncertainty. These velocity measurements are displayed in Figure~\ref{fig:map}b, along with their uncertainties.

The heliocentric radial velocity of the stream can be seen to change smoothly by almost $600\kms$ over the $95\deg$ that we detect it over. We performed a simple empirical fit to the velocity data $v$, rejecting those stars with $|v-v_{\rm fit}|>20\kms + 2 \delta v$, where $\delta v$ is the radial velocity uncertainty. The fitted polynomial
\begin{equation}
v_{\rm fit}=90.68 \, \phi_1^3 + 204.5 \, \phi_1^2 - 254.2 \, \phi_1 - 261.5 
\end{equation}
(with velocities in $\kms$ and $\phi_1$ in radians) is shown with a blue line, and the 117 stars that are retained in the fit are shown in black, while the 39 rejected stars are colored magenta. We deliberately use empirical fits in the present contribution rather than fitting a stream model so as to avoid mismatch biases from errors in the Galactic potential model.

A further empirical fit $S(\phi_1)$ is made to the $\phi_2$ trend of the 117 velocity-confirmed members. We find:
\begin{equation}
S(\phi_1)=0.008367 \, \phi_1^3 - 0.05332 \, \phi_1^2 - 0.07739 \, \phi_1 - 0.02007 \, ,
\end{equation}
where all angles are in radians. This fit is shown with the solid blue line Figure~\ref{fig:map}a. The majority of the velocity-confirmed members lie within $0\degg6$ of this fitted line (i.e., between the dotted lines): discounting the ``spur'' feature, only 4 velocity members extend beyond $0\degg6$. In contrast, of the 39 velocity non-members, 26 lie beyond $0\degg6$. Thus the contamination of the velocity sample within $|S(\phi_1)| < 0\degg6$ is only 11\%, which motivates our choice of selecting stars from the {\tt STREAMFINDER} sample from within this region of sky. 

The proper motion properties of GD-1 are displayed in Figure~\ref{fig:PMs}, derived from the 603 {\tt STREAMFINDER} candidates with $|\phi_2-S(\phi_1)| < 0\degg6$ and that are not radial velocity outliers. The fitted polynomial relations are:
\begin{equation}
\mu_{\alpha, \rm fit}=3.794 \phi_1^3 + 9.467 \phi_1^2 + 1.615 \phi_1 - 7.844
\end{equation}
and
\begin{equation}
\mu_{\delta, \rm fit}=-1.225 \phi_1^3 + 8.313 \phi_1^2 + 18.68 \phi_1 - 3.95 \, ,
\end{equation}
with $\phi_1$ in radians and the proper motions in $\masyr$.

\begin{figure}
\begin{center}
\includegraphics[angle=0, viewport= 1 10 405 390, clip, width=\hsize]{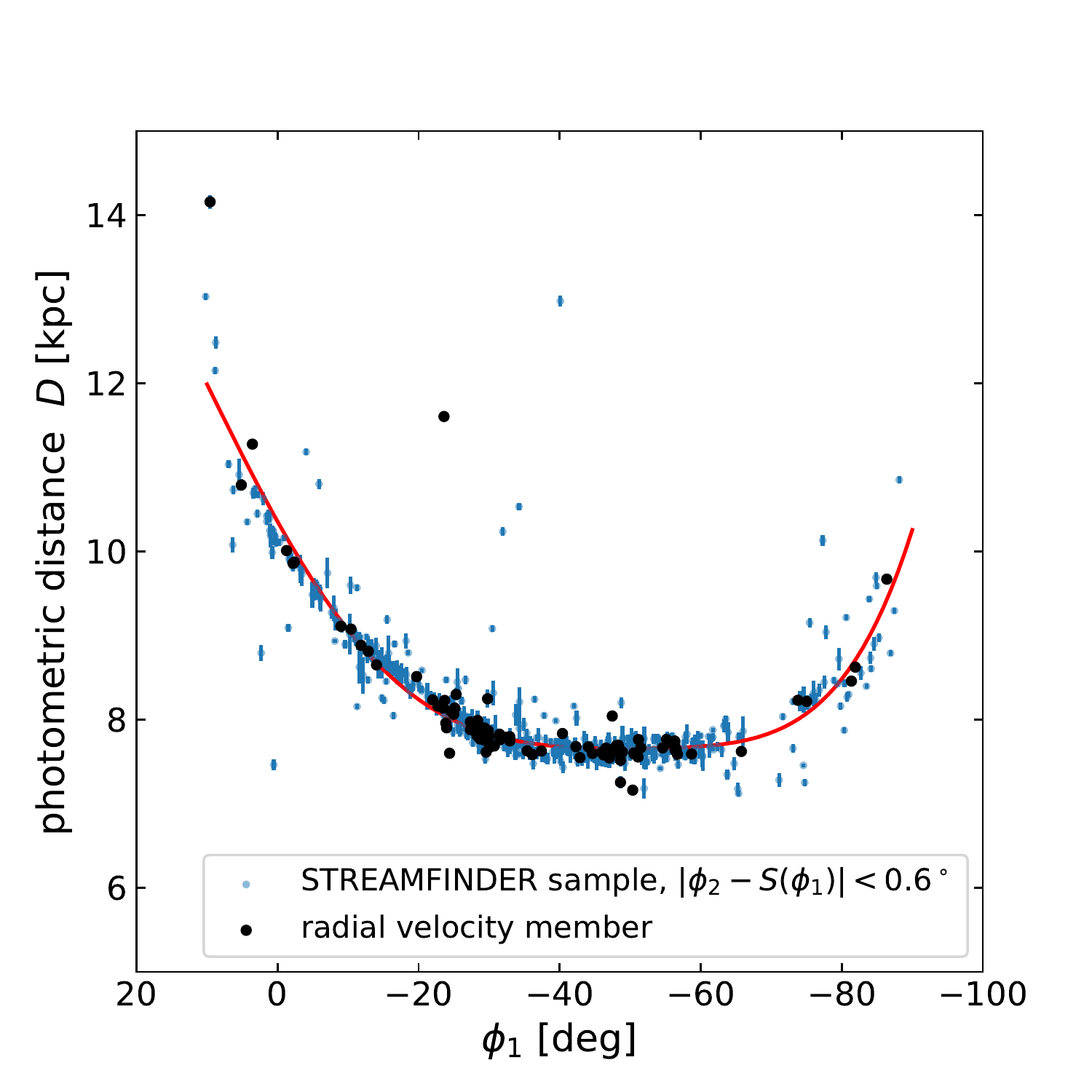}
\end{center}
\caption{Heliocentric distance profile along GD-1, derived by the {\tt STREAMFINDER} using Pan-STARRS photometry in addition to Gaia DR2 data. The red line shows a quintic polynomial fit to the velocity-confirmed data (black dots). (Error bars show $1\sigma$ distance uncertainties, as estimated by the software).}
\label{fig:distance_profile}
\end{figure}

The {\tt STREAMFINDER} software returns the most likely distance solution for the stream model at the position of every star in the sample. The search for stellar streams was initially undertaken using only Gaia photometry, and we conducted our spectroscopic follow-up survey based on those data. However, we have recently updated the software to allow us to include other photometric catalogs. The only conceptual change to the software that this entails is an additional factor in the probability density model of the stream (Equation 2 of \citealt{2019ApJ...872..152I}) to account for the probability of the additional photometric information given the stellar population model prediction. For the present contribution, we have included the Pan-STARRS $g$- and $r$-band photometry, and model its deviation from the PARSEC model predictions with a simple Normal distribution, i.e., $\mathcal{P}_{\rm color, PS1}=\mathcal{N}(x)$, where $x \equiv ((g-r)_0-(g-r)_{0, \rm model})/\delta(g-r)$, where $\delta(g-r)$ is the color uncertainty. This upgrade to the software significantly decreases the uncertainties on the distance estimates. The resulting distance trend is shown in Figure~\ref{fig:distance_profile}, which we have fit (red line) with the following polynomial:
\begin{equation}
\begin{split}
D(\phi_1)=&-4.302 \phi_1^5 - 11.54 \phi_1^4 - 7.161 \phi_3 \\
          &+ 5.985 \phi_1^2 + 8.595 \phi_1 + 10.36 \, ,
\end{split}
\end{equation}
where $D$ is in kpc and $\phi_1$ is in radians. This fit was made to the velocity-confirmed stars (filled black circles), but it clearly also encapsulates the trend of the full {\tt STREAMFINDER} sample (blue points).

The analysis described so far in this section has allowed us to derive empirical fits to the track of the stream on the sky, to its line of sight velocity profile, to the proper motion gradient in $\mu_\alpha$ and $\mu_\delta$, and to the distance gradient. With these ingredients we can now return to the original Gaia catalog and examine the density distribution along the stream.

\begin{figure*}
\begin{center}
\includegraphics[angle=0, viewport= 25 40 670 640, clip, width=\hsize]{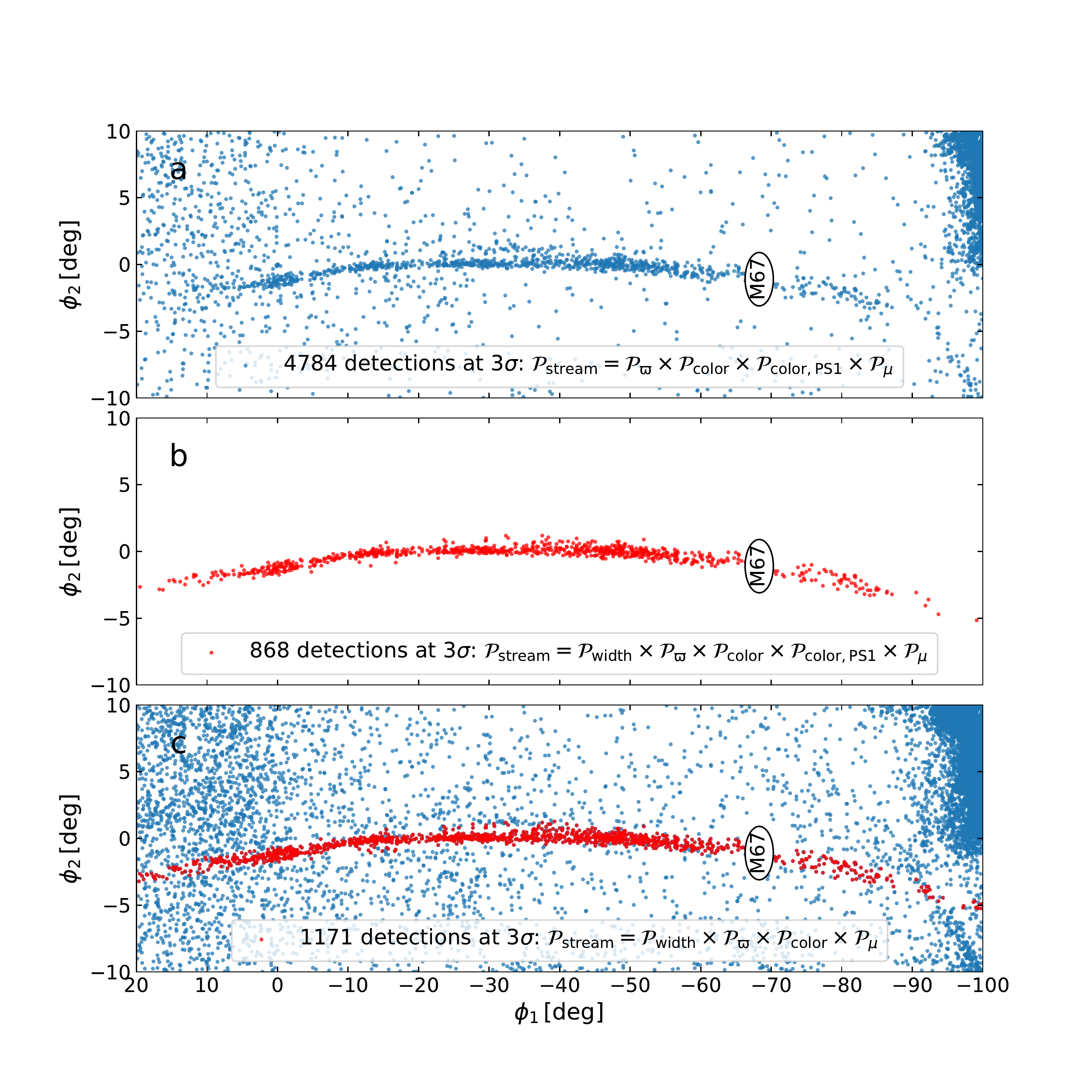}
\end{center}
\caption{Selection of GD-1 candidates directly from the Gaia DR2 and Pan-STARRS DR2 catalogs according to their proximity to the empirical stream model. (a) shows the sources within $3\sigma$ of the model, but ignoring the sky proximity criterion ($\mathcal{P}_{\rm width}$). Moderate contamination can be seen towards the ends of the stream. (b) as (a) but now accounting for the proximity to the  $S(\phi_1)$ track. Using only Gaia photometry (c) leads to somewhat higher contamination.}
\label{fig:Gaia_maps}
\end{figure*}

\section{GD-1 Density Profile}
\label{sec:Density_Profile}

While the {\tt STREAMFINDER} provides a sample of stream members, the reader may be concerned that the algorithm's parameters could bias the results in a complicated way. For this reason, we now proceed to extract two additional stream samples selected in a more traditional way, to serve as comparisons.

We first extract a sample of GD-1 stars from the Gaia DR2 catalog, taking those stars with $G_0<20$~mag, that have a full 5-component astrometric solution, and that possess a flux excess $E (\equiv{\tt phot\_bp\_rp\_excess\_factor})$ in the range:
\begin{equation}
1+0.015 (G_{BP}-G_{RP})^2 < E < 1.3 + 0.06 (G_{BP}-G_{RP})^2 
\end{equation}
(see \citealt{2018A&A...616A...2L} for the motivation for this constraint).

For every star in the Gaia DR2 catalog in the region $-100\deg<\phi_1<20\deg$ and $-10\deg<\phi_2<10\deg$, we calculate the probability of the star belonging to an idealised stream model, which is simply the empirical sky position, distance, and proper motion profiles convolved with appropriate Gaussians:
\begin{equation}
\mathcal{P}_{\rm stream}  = \mathcal{P}_{\rm width} \times \mathcal{P}_{\rm \varpi} \times \mathcal{P}_{\rm color} \times \mathcal{P}_{\rm color, PS1} \times \mathcal{P}_\mu \, ,
\end{equation}
where $\mathcal{P}_{\rm width}$ is the probability that the star is located at the observed $\phi_2-S(\phi_1)$ perpendicular distance from the stream track, $\mathcal{P}_{\rm \varpi}$ is the probability of the observed parallax given the distance model, $\mathcal{P}_{\rm color}$ is the probability of the observed Gaia $G_{BP}-G_{RP}$ color, given the distance model and the stellar population model, and $\mathcal{P}_{\rm color, PS1}$ is the same probability for the Pan-STARRS photometry. These four probability terms are modelled as one-dimensional Gaussians, and the observed uncertainties are taken into account by  adding the uncertainty in quadrature with the intrinsic model dispersion (for the $\mathcal{P}_{\rm \varpi}$ term, we also adopt the parallax zero-point of $-0.029\mas$ found by \citealt{2018A&A...616A...2L}). The fifth factor $\mathcal{P}_\mu$ is the probability of stream membership given the measured proper motion differences from the model ($\Delta_{\mu_\alpha}$ and $\Delta_{\mu_\delta}$), and is given by:
\begin{align}
\begin{split}
\mathcal{P}_\mu & = {{1}\over{2\pi \sigma_{\mu_\alpha} \sigma_{\mu_\delta} \sqrt{1-\rho^2}}}  \times \\
& \exp(-{{1}\over{2(1-\rho^2)}} \Bigg[ {{\Delta^2_{\mu_\alpha}}\over{\sigma^2_{\mu_\alpha}}} + {{\Delta^2_{\mu_\delta}}\over{\sigma^2_{\mu_\delta}}}  
- {{2\rho \Delta_{\mu_\alpha} \Delta_{\mu_\delta}}\over{\sigma_{\mu_\alpha} \sigma_{\mu_\delta}   }}  \Bigg] )  \, .
\end{split}
\end{align}
We thus take into account the proper motion uncertainties $\sigma_{\mu_\alpha}$ and $\sigma_{\mu_\delta}$ their correlation $C \equiv {\tt pmra\_pmdec\_corr}$ (see \citealt{2018A&A...616A...2L}), which is incorporated into the term
\begin{equation}
\rho = {{ C \, \sigma_{\mu_\alpha} \sigma_{\mu_\delta}}\over {\sqrt{ (\sigma^2_{\mu_\alpha} + w^2_\mu) (\sigma^2_{\mu_\delta} + w_\mu^2)}}} \, ,
\end{equation}
which can be derived by convolving the two-dimensional covariance matrix with an isotropic two-dimensional Gaussian of dispersion $w_\mu$.

\begin{figure}
\begin{center}
\includegraphics[angle=0, viewport= 15 90 540 1155, clip, width=\hsize]{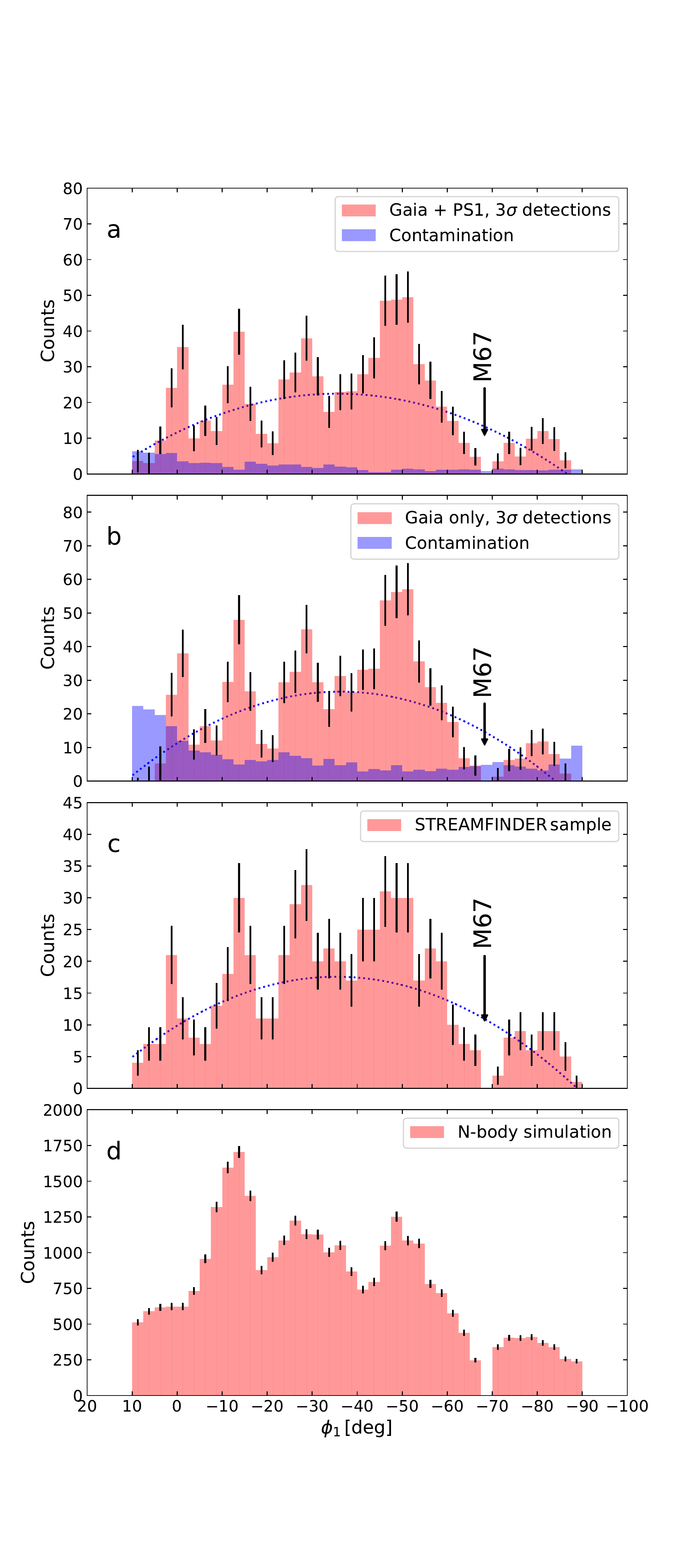}
\end{center}
\caption{Counts (with $1\sigma$ uncertainties) of candidate members as a function of $\phi_1$. (a) shows the density profile based on Gaia data, complemented with Pan-STARRS photometry, based on the map in Figure~\ref{fig:Gaia_maps}a. The blue histogram shows the expected contamination level derived from the sky region at $|\phi_2-S(\phi_1)|>2\deg$. The quadratic fit (dotted line) is defined in the text. (b) shows the same information as (a), but ignoring the PS1 information. (c) displays the profile derived from the {\tt STREAMFINDER} detections with $|\phi_2-S(\phi_1)|<0\degg6$, where the radial velocity non-member stars identified in Figure~\ref{fig:map} have been rejected. (d) shows the profile of the N-body simulation presented in  Section~\ref{sec:Modelling_the_Density_Profile}, which possesses some of the main features seen in the observations. }
\label{fig:spikes}
\end{figure}

\begin{figure*}
\begin{center}
\includegraphics[angle=0, viewport= 25 1 670 325, clip, width=\hsize]{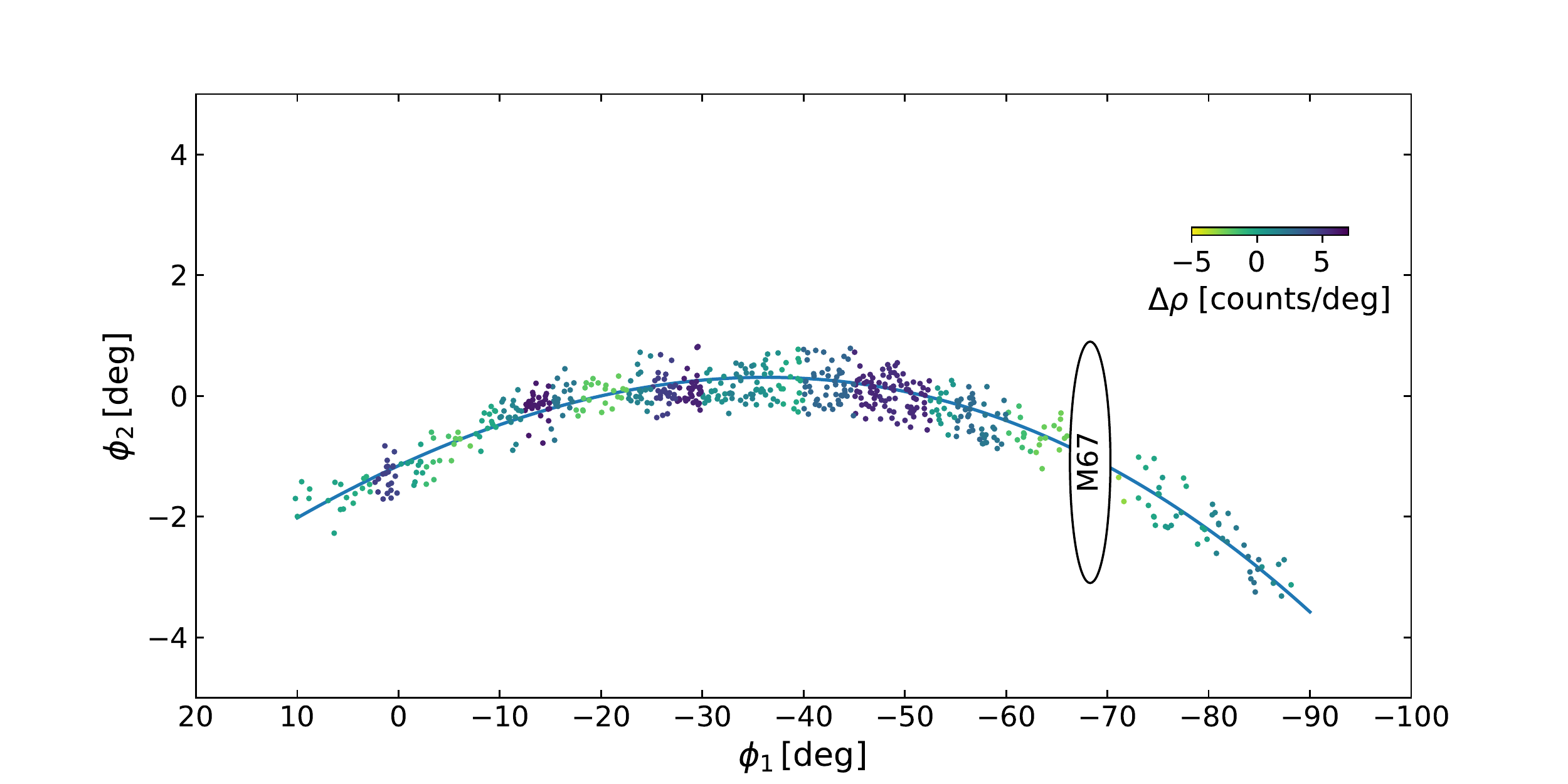}
\end{center}
\caption{Map of the {\tt STREAMFINDER} GD-1 sources color-coded by local density enhancement over a quadratic fit to the counts shown in Figure~\ref{fig:spikes}c.}
\label{fig:map_density}
\end{figure*}

For the present search, we assume a Gaussian width of the stream of $50\pc$, and we (generously) allow a dispersion in proper motion $w_\mu$ equivalent to $10\kms$ in velocity. We adopt the same three PARSEC stellar populations models as used above for the {\tt STREAMFINDER}, with age and metallicity ($12.5\Gyr, -2.0$), ($12.0\Gyr, -2.3$), ($10.0\Gyr, -1.7$), and for each star, we select the solution that yields the highest probability. 

The result of the search is shown in Figure~\ref{fig:Gaia_maps}. In panel (a) we show the 4784 stars that lie within $3\sigma$ of the model in parallax, proper motion, and Gaia and Pan-STARRS photometry. While GD-1 is clearly visible, there is a non-negligible amount of contamination in the map, especially towards the extremities of the structure, where it approaches regions of low Galactic latitude. Panel (b) also shows a $3\sigma$ membership cut, but this time the distance to the stream track is also taken into account, yielding a sample of 868 stars. A comparison to (a) shows that this $3\sigma$ cut corresponds to the sample that one would select visually as probable GD-1 members. Note that this $3\sigma$ cut removes the ``spur'' feature \citep{2018ApJ...863L..20P} visible in Figure~\ref{fig:map}. Since we are attempting to ascertain the reliability of the density profile, it is useful to consider the properties of alternative GD-1 samples. Therefore, in panel (c), we present the map of the 1171 stars that lie within $3\sigma$ of the empirical model, but this time ignoring the Pan-STARRS photometry. As expected, the contamination is higher in this case. 

Figure~\ref{fig:spikes} condenses this information into one-dimensional star-count profiles, where the $3\sigma$ sample derived with Gaia and Pan-STARRS information is presented in (a), while (b) ignores the Pan-STARRS colors. The estimated contamination in each sample is shown in the blue histograms, and has been subtracted from the density profiles of interest (red histograms). This contamination is estimated by selecting those stars with $|\phi_2-S(\phi_1)|>2\deg$ and with $|\phi_2|<10\deg$. In (c) we show the profile of the {\tt STREAMFINDER} sample of 603 stars within $|\phi_2-S(\phi_1)|<0\degg6$ (and that are not radial velocity outliers). A simple quadratic was fitted to each profile (dotted lines); these are, respectively, for panels (a,b,c):
\begin{equation}
    \begin{split}
        C_{\rm Gaia+PS1}(\phi_1)=&-37.51 \phi_1^2 - 46.51 \phi_1 + 14.37\\
        C_{\rm Gaia}(\phi_1)=&-50.48 \phi_1^2 - 63.37 \phi_1 + 13.86\\
        C_{\tt STREAMFINDER}(\phi_1)=&-28.7 \phi_1^2 - 35.96 \phi_1 + 11.26 \, ,
    \end{split}
\end{equation}
where $\phi_1$ is in radians and the counts $C$ are per bin of width $2\degg5$. Interestingly, the density distribution along the stream displays prominent spikes that can be seen as high contrast peaks above the low-order fit. In Figure~\ref{fig:map_density} we reproduce the sky distribution of these sources in the {\tt STREAMFINDER} sample, colored according to local density.

It is not surprising that the three distributions in Figure~\ref{fig:spikes}a--c are not identical given the different selection procedures. However, they all display at least four extremely prominent peaks at the same locations, and as such they paint a consistent picture of the large-scale properties of the GD-1 system.

\begin{figure}
\begin{center}
\includegraphics[angle=0, viewport= 10 10 400 390, clip, width=\hsize]{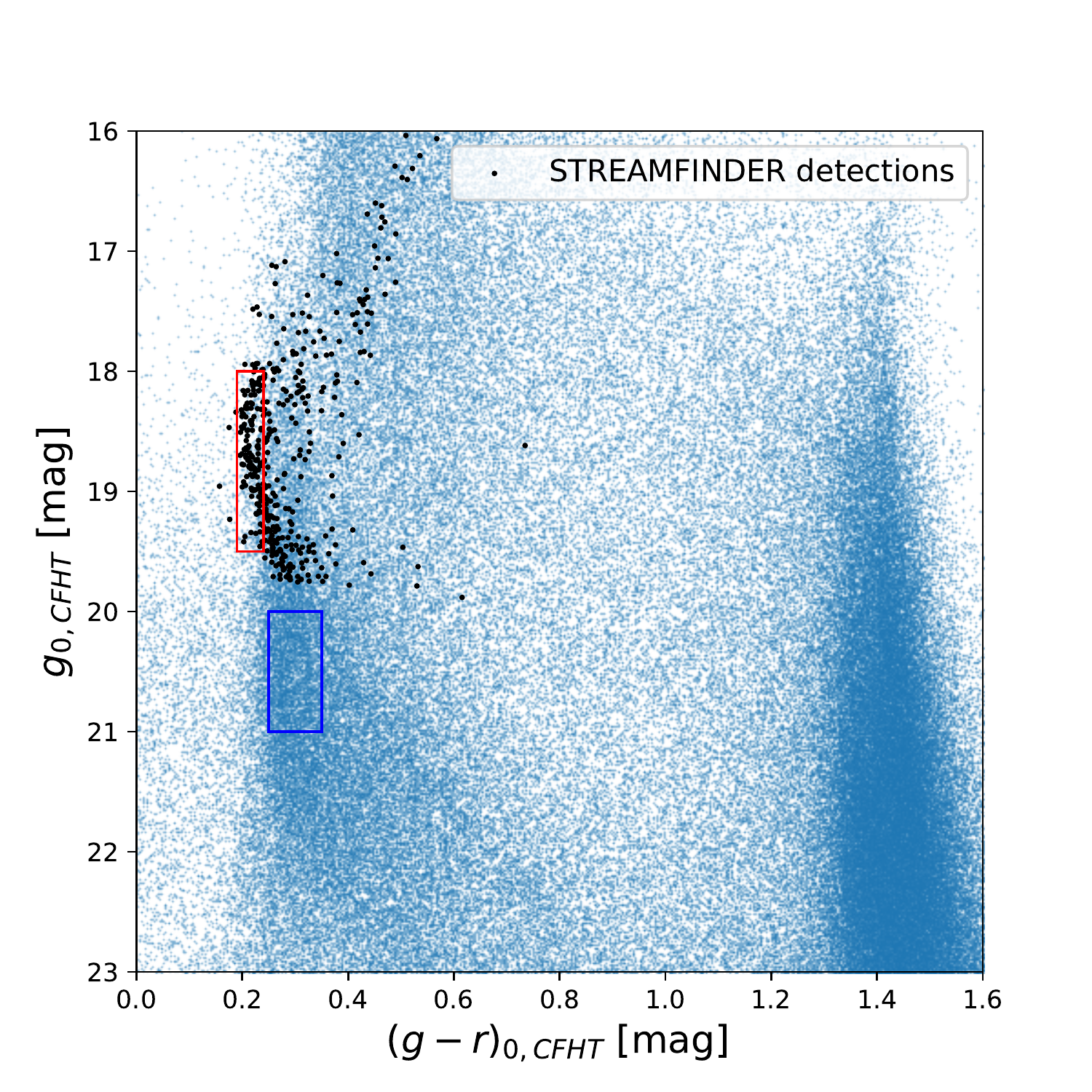}
\end{center}
\caption{Color-magnitude distribution of sources in the CFHT/MegaCam survey (blue dots). The stars identified as candidate members in this region by the {\tt STREAMFINDER} algorithm are highlighted with larger black dots. These can be seen to follow a color-magnitude sequence that includes the main sequence turnoff (red rectangle), where the GD-1 population has the highest contrast over the Galactic contamination. The blue rectangle selects a small dense portion of the main sequence turn-off of the halo.}
\label{fig:CMD}
\end{figure}

\begin{figure}
\begin{center}
\includegraphics[angle=0, viewport= 10 10 400 390, clip, width=\hsize]{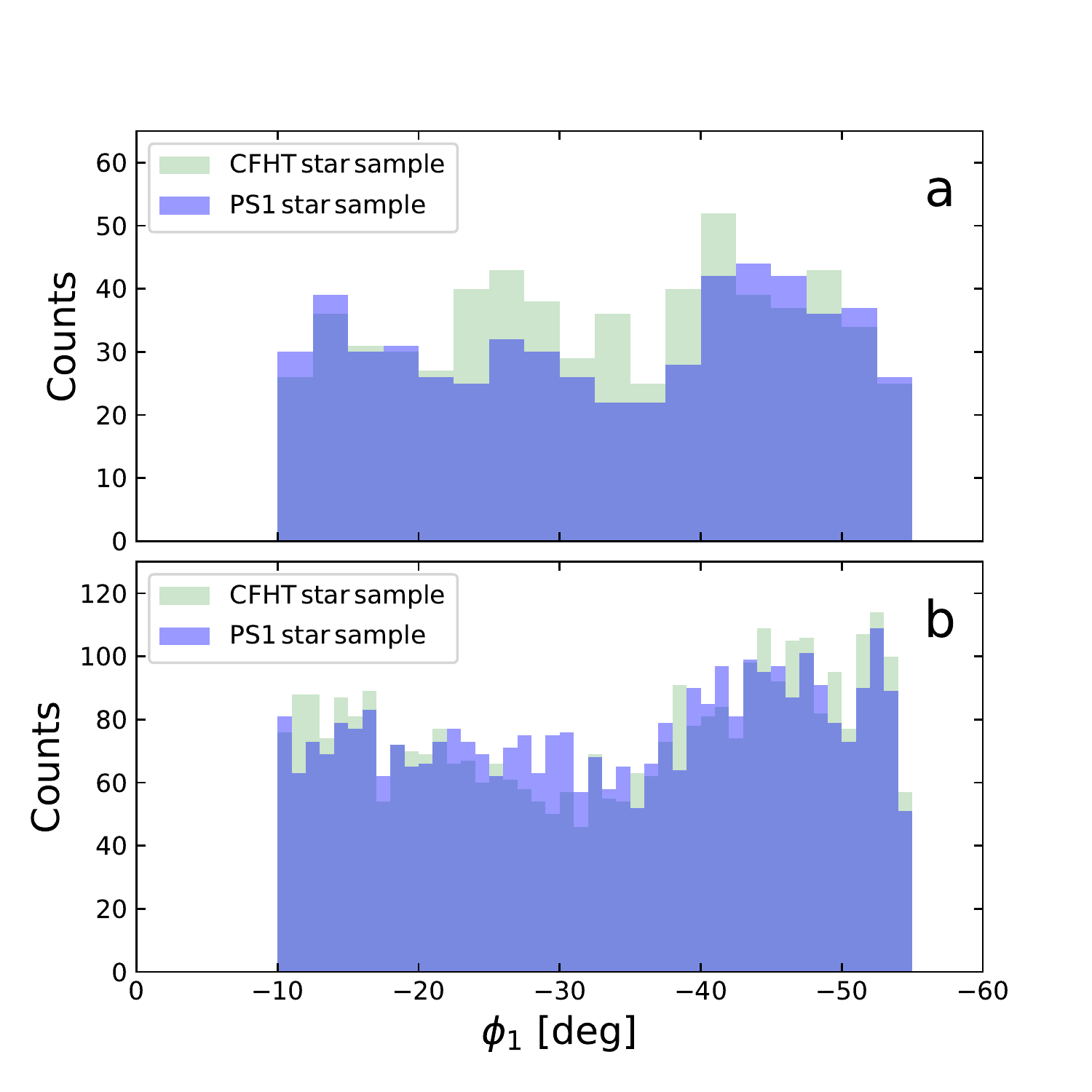}
\end{center}
\caption{a: Counts as a function of $\phi_1$ of stars within the red selection box of Figure~\ref{fig:CMD}. The CFHT/MegaCam sample (green) is compared to the Pan-STARRS sample (blue), both selected with identical color-magnitude criteria (the Pan-STARRS bands are converted to CFHT colors, as explained in the text). We have intentionally omitted the Poisson uncertainties on these distributions, so as to highlight the fact that they are not independent samples. In principle the distributions should be identical; but they are not due to differences in completeness and star/galaxy classification between the two surveys. The substantial ($\sim 20$\%) variations will result in large spatially-dependent errors in any density map derived from such data. This problem is further compounded if the observed counts are dominated by a large contaminating population that needs to be subtracted off (as is the case with GD-1). b: deviations between the counts of point-sources selected within the blue rectangle in Figure~\ref{fig:CMD}.}
\label{fig:Counts_comparison}
\end{figure}

\section{Density artifacts due to mis-classification and incompleteness}
\label{sec:CFHT_Data}

Before analysing the implications of our measurement of the density profile along GD-1, we will first discuss the limitations of stellar density maps derived from ground-based imaging. Modern wide-field cameras allow one to detect structures of very low surface brightness simply by counting individual resolved stars. Typically, these sources are revealed as small enhancements over the foreground and background contaminating populations. Generally, the image quality of a camera degrades away from the field center, causing both higher photometric uncertainties, and poorer classification constraints, so that the fraction of stars that may be confused with galaxies (and vice-versa) worsens towards the edges of the field of view.

The weather conditions obviously also change over the course of a large survey, leading to varying survey depths as the transparency of the sky changes, as well as different depths for accurate star/galaxy classification. Temperature variations will also lead to variations in the quality of the focus. 

All these factors affect the spatial homogeneity of a survey in a complicated way that is not easy to estimate or correct for. This is especially the case for public surveys where the information about the observing conditions that went into producing the data in a particular region of sky are difficult to recover. We therefore felt that it would be useful to investigate how reliably a ground-based survey such as Pan-STARRS could be used to measure large-scale stellar density. We stress that Pan-STARRS photometry is known to be photometrically extremely well calibrated (with a reliability of 7--12 millimags; \citealt{2016arXiv161205560C}); the issue we wish to assess here is its homogeneity to classification and completeness over large fields.

To this end, we decided to compare the GD-1 stream region in Pan-STARRS to a deeper survey, taken in good seeing conditions with CFHT/MegaCam, which was previously analysed by \citet{2018MNRAS.477.1893D}. We retrieved the images from the CFHT archive and processed them with the same procedure as applied to data from the Canada-France Imaging Survey \citep{2017ApJ...848..128I}. The dataset consists of 528 $g$-band images and 516 $r$-band images, all of exposure time $50$~s, that cover the gray-shaded region in Figure~\ref{fig:map}a.

The CFHT/MegaCam images were recalibrated onto the Gaia DR2 astrometric reference, which was also used as the astrometric reference for the Pan-STARRS DR2 catalog. The zero-points of the CFHT/MegaCam $g$ and $r$-band photometry were calibrated onto the Pan-STARRS DR2 survey, adopting the color transformations\footnote{see \url{http://www.cadc-ccda.hia-iha.nrc-cnrc.gc.ca/en/megapipe/docs/filt.html}}:
\begin{equation}
\begin{split}
g_{\rm CFHT}&=g_{\rm PS1}+0.014+0.059 x - 0.00313 x^2 - 0.00178 x^3\\ 
r_{\rm CFHT}&=r_{\rm PS1}+0.003-0.050 x + 0.0125  x^2 - 0.00699 x^3
\end{split}
\end{equation}
where $x\equiv (g-i)_{\rm PS1}$. The Cambridge Astronomical Survey Unit (CASU) software \citep{2001NewAR..45..105I} was used to measure the photometry and perform the star/galaxy classification.

Figure~\ref{fig:CMD} shows (in blue) the resulting color-magnitude diagram (CMD) of all stellar sources identified in the CFHT/MegaCam survey. The black points mark the positions of the {\tt STREAMFINDER} GD-1 candidates that are present in the CFHT/MegaCam survey region; the main sequence turnoff of GD-1 is clearly visible, and the sample also contains some sub-giant and red giant branch stars.

\begin{figure*}
\begin{center}
\includegraphics[angle=0, viewport= 55 10 785 375, clip, width=\hsize]{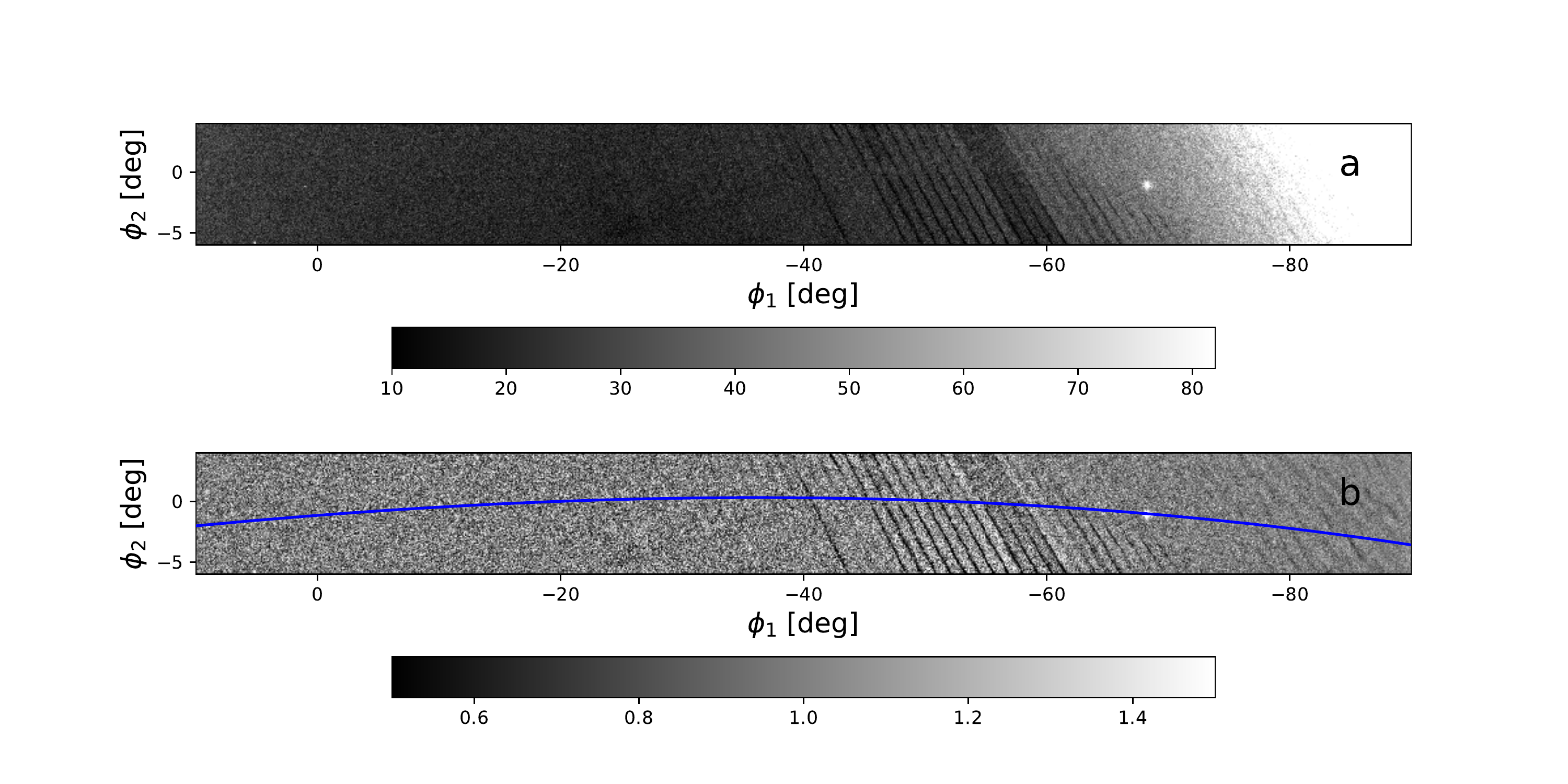}
\end{center}
\caption{Spatial artifacts in Gaia DR2 in the vicinity of GD-1. Panel (a) shows the density of all stars to $G_0=20$~mag in pixels of $0\degg1\times0\degg1$. The corresponding count level is displayed below the image. The diagonal dark stripes are regions with a lower number of observations due to Gaia's scanning pattern. In (b) we present a ``flat-field'' map, constructed by dividing the distribution in (a) by a smoothed version of itself (using a Gaussian kernel with $\sigma=1\deg$). The track of GD-1 in these coordinates  (blue) is reproduced from Figure~\ref{fig:map}a.}
\label{fig:Gaia_artifacts}
\end{figure*}

For our comparison test, we decided to isolate the stars in the red selection box in Figure~\ref{fig:CMD}, as this corresponds to the CMD location where GD-1 has its highest contrast over the contaminating populations of the Milky Way, and it will be the signal in this CMD region that a matched filter will enhance. A rectangular box is chosen for simplicity, selecting stars with $(g-r)_0\in[0.19,0.24]$ and $g_0\in[18,19.5]$. The average photometric uncertainties of the stars in this box are below $0.01$~mag in both surveys and in both colors. A sample of point-sources is selected from the Pan-STARRS DR2 survey by (conservatively) retaining only those sources where the $r$-band PSF magnitudes agree with the aperture magnitudes to within $0.05$~mag. The Pan-STARRS targets are further required to have a minimum of two detections in the $g$- and $r$-bands, and to have ${\tt qualityFlag}=4$ (which identifies good-quality measurements in Pan-STARRS).

\begin{figure}
\begin{center}
\includegraphics[angle=0, viewport= 5 10 400 390, clip, width=\hsize]{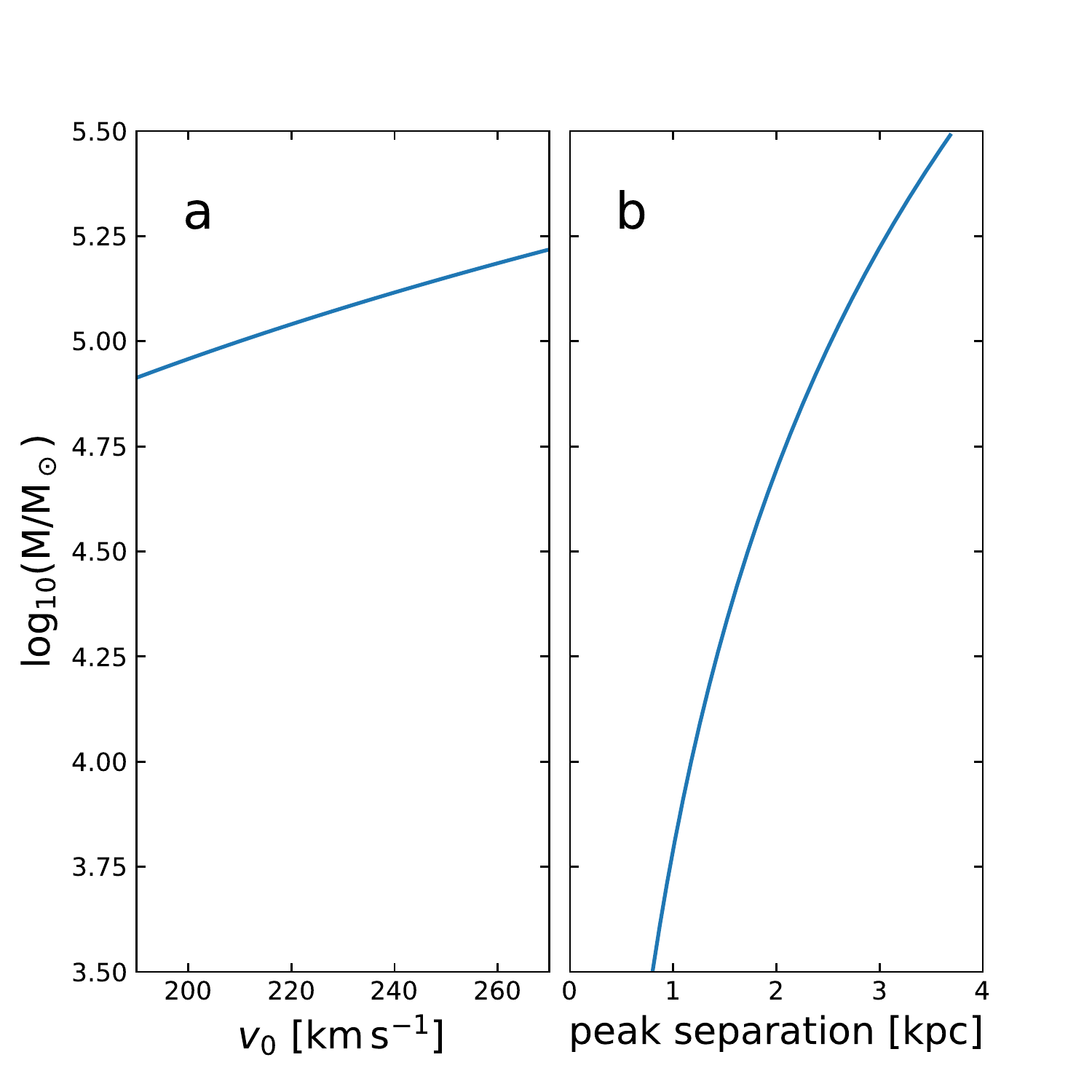}
\end{center}
\caption{(a) Relation between the circular velocity of the Milky Way $v_0$ and the mass of the GD-1 cluster $M$, see Eq.~(\ref{eq:Mv0rel}). (b) Separation of the density peaks as a function of the satellite mass, predicted from Eq.~(\ref{eq:yC}), assuming a flat circular velocity curve where $v_0=229\kms$.}
\label{fig:analytic}
\end{figure}

Figure~\ref{fig:Counts_comparison}a shows a comparison of the counts in the CMD selection box along the length of the stream section where the CFHT/MegaCam imaging was obtained. Substantial $\sim 20$\% variations are seen in what one might have supposed to have been almost identical overlapping samples. Figure~\ref{fig:Counts_comparison}b repeats this test, but for a slightly fainter sample (selected within the blue rectangle in Figure~\ref{fig:CMD} with $(g-r)_0\in[0.25,0.35]$ and $g_0\in[20,21]$), which contains halo main sequence turn-off stars. Given that the halo is expected to be an ancient, dynamically well-mixed population, this sample should be spatially smooth. Within this box the typical uncertainties of the CFHT/MegaCam and Pan-STARRS photometry are 0.01~mag and 0.05~mag, respectively. Significant differences in the number-counts profiles between the CFHT/MegaCam and Pan-STARRS results are seen again, and a comparison between panels a and b of Figure~\ref{fig:Counts_comparison} demonstrates that the deviations do not match up spatially between the samples.

In addition to the large-scale variations of the type seen in Figure~\ref{fig:Counts_comparison}a, which may be due to variable transparency and seeing over the course of a survey, periodic camera-sized density variations are often seen in wide-field maps. Such artifacts can sometimes be spotted following the survey tiling pattern (see, e.g., the ripples in the PAndAS survey in Figure~11 of \citealt{2007ApJ...671.1591I}, or in the $u$-band of the SDSS in Figure~3 of \citealt{2017ApJ...848..128I}).

The astrophysical interpretation of density variations measured from ground-based wide-field surveys therefore requires a very careful correction for spurious signals.

Of course space-missions may also have spatially-dependent artifacts. In the case of Gaia DR2 there are particularly noticeable stripes of incompleteness that follow the scanning pattern (see e.g. \citealt{2018A&A...616A...2L}), and these artifacts will contribute to the measured spatial variations in density. These problems are difficult to perceive in our previous maps, because of the low density of sources in the stream. However, by examining the spatial distribution of all Gaia sources with $G_0<20$~mag (Figure~\ref{fig:Gaia_artifacts}) sufficient statistics are attained to reveal numerous track-like diagonal under-densities crossing the path of GD-1. These are particularly noticeable in the interval $\phi_1=[-60\deg,-40\deg]$, where they cause narrow ($\sim 0\degg2$) dips of $\sim 50$\% lower density with a periodicity in $\phi_1$ of slightly over $1\deg$. A wider ($\sim 3\degg6$) band of lower density is also visible intersecting the stream path at $\phi_1=-56\degg5$. These artifacts cause a patchy incompleteness in the stream survey, and will contribute spurious gap-like information to the density power spectrum of GD-1 derived from Gaia DR2 data.

\begin{figure*}
\begin{center}
\includegraphics[angle=0, viewport= 44 3 610 335, clip, width=13cm]{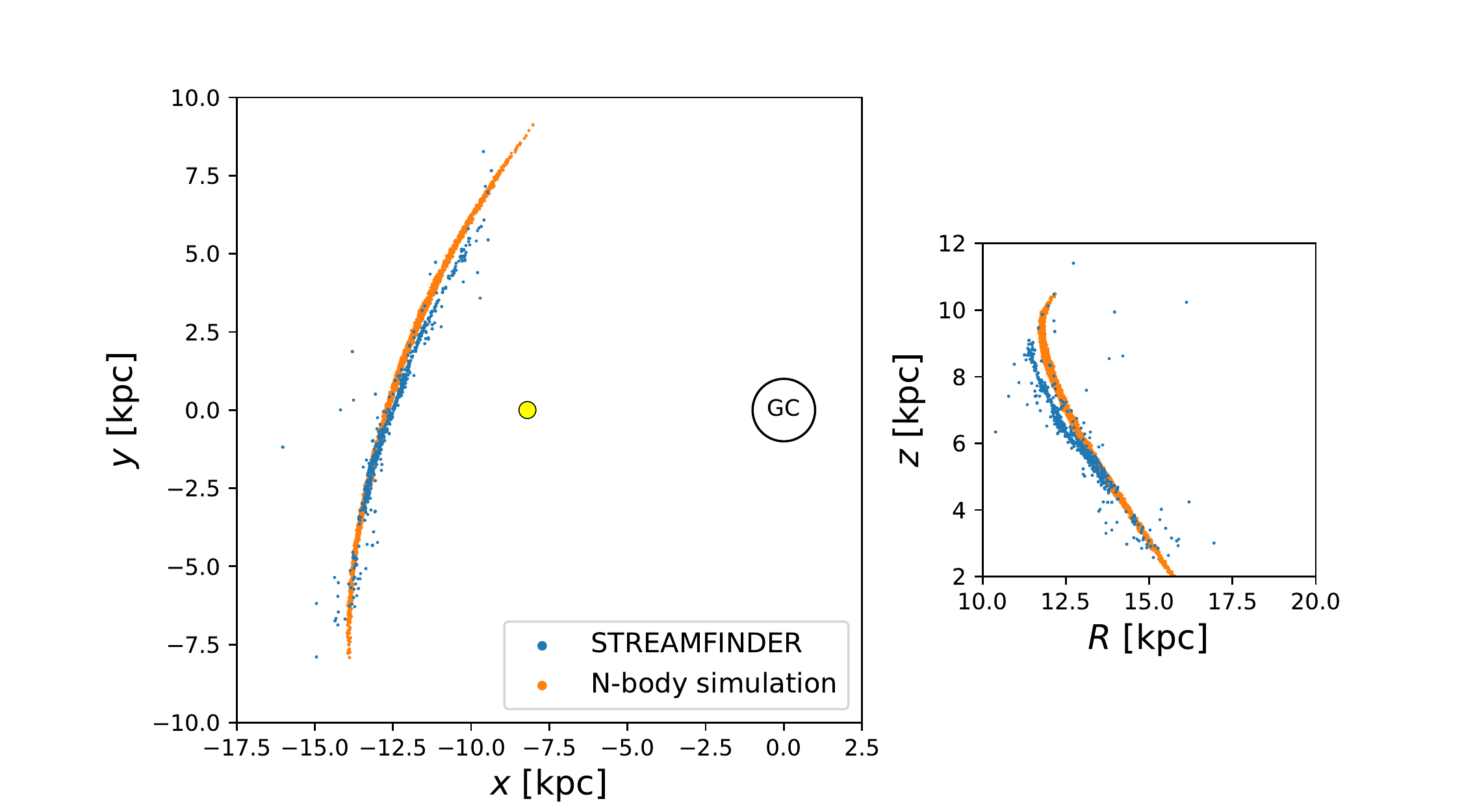}
\end{center}
\caption{Positions of N-body model particles (orange) are compared to positions of the GD-1 stars (blue), as as derived using the {\tt STREAMFINDER} distances shown in Figure~\ref{fig:distance_profile}.}
\label{fig:positions3D}
\end{figure*}

\section{Modelling the Density Profile}
\label{sec:Modelling_the_Density_Profile}

Having presented some of our concerns on the limitations of star-counts measurements, we now proceed to model the density profiles measured in Section~\ref{sec:Density_Profile}. The star distribution of Figure~\ref{fig:map_density} is strikingly reminiscent of the epicyclic overdensities seen in simulations of slowly disrupting clusters: see especially Figure~7 of \citet{2012MNRAS.420.2700K}. This suggests that we can constrain the mass of the progenitor from the periodicity of the density peaks. We consider here the simplest case, in which a satellite moves on a circular orbit, and material is lost at the escape radius with null velocity. Given that GD-1 lies on a low eccentricity orbit ($e=0.33$, \citealt{2009ApJ...697..207W}), this simple configuration is not too unrealistic. In this situation the distance between two overdensities due to the epicycles along the stream is \citep{2008MNRAS.387.1248K}
\begin{equation}\label{eq:yC}
    \yC = - \frac{4\pi\Omega}{\kappa}\left(1-\frac{4\Omega^2}{\kappa^2}\right)\xE,
\end{equation}
where $\Omega$ and $\kappa$ are the circular and epicyclic frequency at the Galactocentric distance $r$ of the cluster and $\xE$ is the escape radius from the cluster. The escape radius can be approximated by 2.88 times the Jacobi radius \citep{2011MNRAS.417..198V,2015MNRAS.452..301F}
\begin{equation}\label{eq:xL}
    \xE = 2.88 \times \left(\frac{GM}{4\Omega^2-\kappa^2}\right)^{1/3} \, ,
\end{equation}
where $M$ is the total mass of the cluster.

If we assume a spherical Milky Way model with a constant circular velocity $v_0$ (i.e. with a logarithmic potential), then
$\Omega(r) = v_0/r$, and $\kappa(r)=\sqrt{2}\Omega(r)$. Eqs.~(\ref{eq:yC}) and (\ref{eq:xL}) then simplify to
\begin{equation}\label{eq:yClog}
    \yC = \frac{4\pi}{\sqrt{2}}\xE,
\end{equation}
and
\begin{equation}\label{eq:xLlog}
    \xE = 2.88 \times \left(\frac{GMr^2}{2v_0^2}\right)^{1/3} \, ,
\end{equation}
respectively. Assuming that we are observing a distance between the peaks along the stream of $\yobs$, this implies a relationship between $v_0$ and $M$
\begin{equation}\label{eq:Mv0rel}
    M \simeq \frac{\yobs^3 v_0^2}{270.26 \times G\pi^3 r^2} \, .
\end{equation}
Inspection of Figure~\ref{fig:distance_profile} reveals that the distance of the stream in the zone of interest is $d\sim 8\kpc$, and the angular distance of the peaks $\Delta\phi \sim 15\deg$. This would suggest that $\yobs = d \Delta\phi \sim 2\kpc$, but this estimate ignores the fact that the stream is not perpendicular to the line of sight. After correcting for the projection effect, we will measure $\yobs = 2.64\pm0.18\kpc$ in Section~\ref{sec:Power_Spectrum_Analysis}.
Given that the stream lies at a Galactocentric distance $r\sim 15\kpc$, we then expect the relationship between circular velocity and progenitor mass shown in Figure~\ref{fig:analytic}a. 

In Figure~\ref{fig:analytic}b, we show the separation of the density peaks of GD-1, predicted from Eq.~(\ref{eq:yC}), taking a circular velocity curve with $v_0=229\kms$ (consistent with the measurement of $v_0=229.0\pm{0.2}\kms$ at the Solar radius by \citealt{2019ApJ...871..120E}), and using a distance of GD-1 from the Sun of $d=8\kpc$.

Having established plausible masses for the GD-1 progenitor, we now examine whether a disrupting N-body model can give rise to the observed stream density profile. For this N-body simulation, we adopted the Galactic potential of \citet{1998MNRAS.294..429D} (their model 1) for the bulge, thin disk, thick disk and interstellar medium. For the dark matter halo, we used a \citet{1997ApJ...490..493N} model similar to the dark matter halo found recently by \citet{2019arXiv191104557C}, with a virial radius of $206\kpc$, a concentration of $c=12$, but with an oblateness of $q=0.82$ (as derived by \citealt{2019MNRAS.486.2995M} from modelling GD-1). These choices lead to a dark halo mass of $9.6\times 10^{11} \msun$. With this Galactic potential model, the circular velocity at the Solar radius ($R_\odot=8.129\kpc$, \citealt{2018A&A...615L..15G}) is $229\kms$. We integrated backwards in time for $2\Gyr$ starting from ${\rm (R.A.,Dec)}=(157\degg6, \, 43\degg71667)$, $d=8.25\kpc$, $(\mu_\alpha,\mu_\delta)=(-6.53 \masyr, -11.0 \masyr)$, and $v_{\rm helio}=-90\kms$. We then integrated a King-model \citep{1966AJ.....71...64K} forwards in time for $2\Gyr$, using the {\tt gyrfalcON} N-body integrator \citep{2000ApJ...536L..39D}. 

The King model was set up to produce a rapidly-disrupting structure so that at the end-point of the simulation there would be no discernible bound structure. The initial mass of the best model we found is $3\times 10^4\msun$ (a factor of $\sim 3$ lower than our prediction from Figure~\ref{fig:analytic}). The model also possesses a central potential $W=3.0$ and a King model tidal radius of $r_t=0.17\kpc$. We used $50,000$ particles, and a softening length of $1.5\pc$. 

In order to account for the incompleteness of the Gaia DR2 survey, we applied the Gaia completeness ``flat field'' map (shown in Figure~\ref{fig:Gaia_artifacts}b, additionally excising a $2\deg$ circle around M67) to the final simulation output.

At the end of the simulation the spatial structure of the stream follows the derived large-scale three-dimensional properties of GD-1 fairly well, as we show in Figure~\ref{fig:positions3D}. Although the cluster is completely disrupted by this point (as observed), we estimate that the position of the progenitor in this model if it had survived would have been $\phi_1= -29\degg925$, $\phi_2= 0\degg096$. Several strong  peaks can be seen to be present in the density profile, as shown in Figure~\ref{fig:spikes}d. Such peaks are a generic property of simulated streams that dissolve slowly in this way \citep{2012MNRAS.420.2700K}. 

Our limited exploration of the parameter space of the simulations suggests to us that it is challenging to match an N-body stream model to these observations, in part because of the cubic dependency of the tidal radius on the progenitor's mass (Eq.~\ref{eq:xL}), and because of the rapid time evolution from what must have been a bound structure to complete dissolution. This renders the location and contrast of the peaks in the N-body simulation very sensitive to the modelled initial conditions. 

Despite displaying a multi-peaked density profile, the best N-body model we have found so far does not reproduce faithfully the observed peak morphology (Figure~\ref{fig:spikes}d). It is particularly noticeable that the peaks are wider than in reality, and the narrow peak at $\phi_1\sim 0\deg$ is not present. Some of our N-body models do produce the density spike at $\phi_1\sim 0\deg$, but obtaining that peak comes at the cost of much lower peak contrast elsewhere in the profile. We are currently in the process of simulating a large library of such models, which will be presented in a future contribution. It is likely that the slight discrepancies between the path of the stream through the Galaxy in the simulation and in the observations that can be seen in Figure~\ref{fig:positions3D} are due to the adopted Galactic potential model giving a slightly incorrect acceleration field; this also will be explored in future work.

We suspect that the large (factor of $\sim 3$) overestimate of the progenitor's mass made by the analytic model (Eq.~\ref{eq:xLlog}) compared to our best N-body simulation is because the assumptions underlying that model do not hold true. In particular, the assumption of constant mass is obviously a poor one in relation to a structure that ends up disintegrating completely.

\section{Power Spectrum Analysis}
\label{sec:Power_Spectrum_Analysis}

\begin{figure*}
\begin{center}
\includegraphics[angle=0, viewport= 15 5 855 425, clip, width=\hsize]{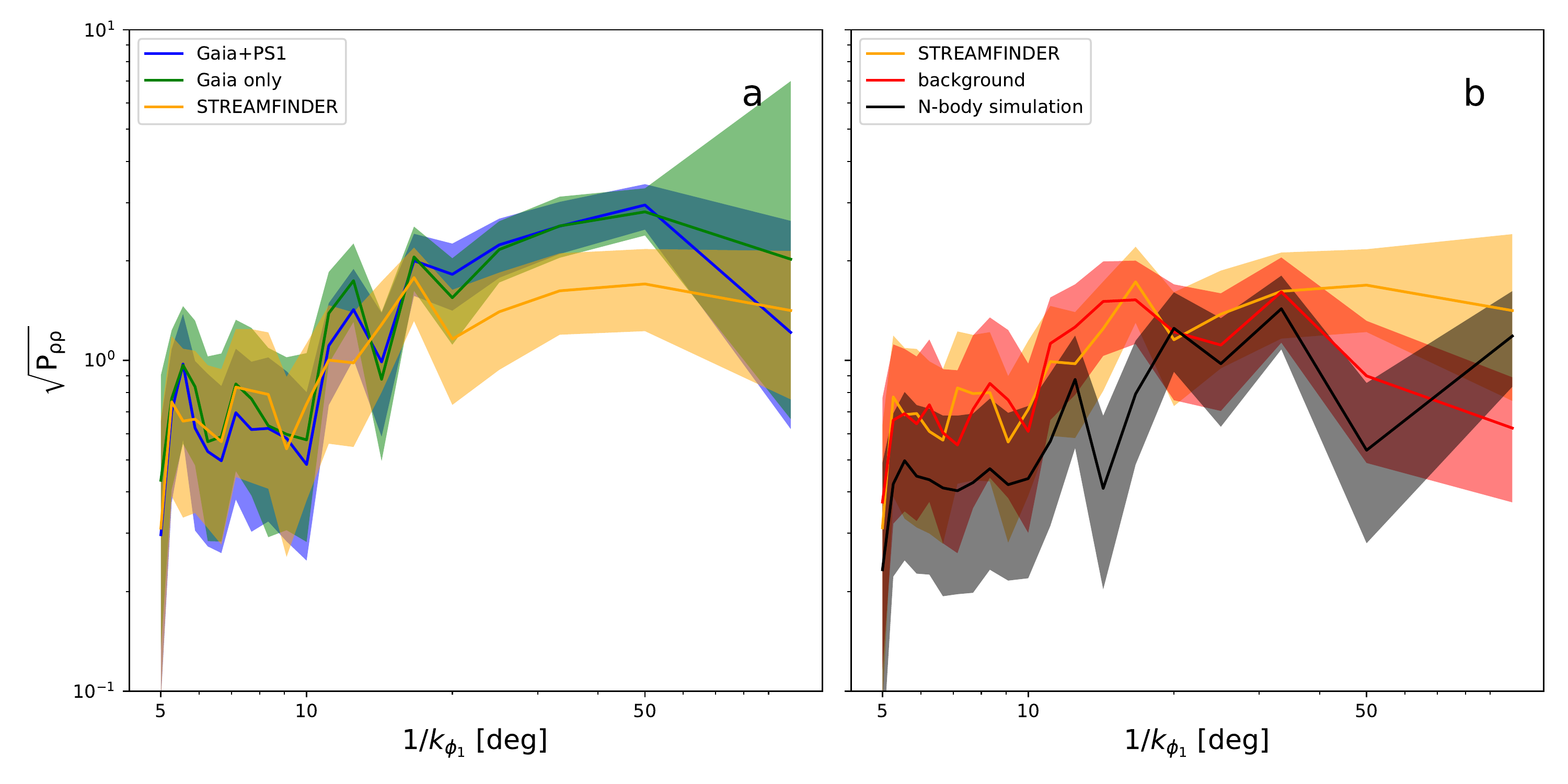}
\end{center}
\caption{Density power spectra of the stream density profiles from Figure~\ref{fig:spikes}. These are calculated using the \citet{1967ITAE...15...70W} method, as provided by the {\tt csd} algorithm in {\tt scipy}. The ($1\sigma$) uncertainties on each power spectrum have been estimated by re-sampling the corresponding density profiles 1000 times. For clarity, the power spectra have been separated into two panels, and the results for the {\tt STREAMFINDER} sample are reproduced in both panels to allow easier comparison. The power spectrum of the background sample (red) corresponds to the contamination profile estimated for the Gaia+PS1 sample. The black line is derived for the N-body model presented in Section~\ref{sec:Modelling_the_Density_Profile}.}
\label{fig:powerspectrum}
\end{figure*}

For completeness, we finally calculate the power spectrum of the tidal stream following \citet{2019arXiv191102662B}. However, we feel that at present the sky position of the remnant of the GD-1 progenitor remains highly conjectural \citep{2018ApJ...863L..20P,2018MNRAS.478.3862M,2018MNRAS.477.1893D,2019arXiv191105745D,2019MNRAS.485.5929W}, so splitting the structure into leading and trailing arms is not justified. We use the {\tt csd} algorithm in {\tt scipy} to calculate the density power spectrum of the profiles shown in Figure~\ref{fig:spikes}a--c normalised by the respective quadratic fits to the continuum. The result of this calculation is shown in Figure~\ref{fig:powerspectrum}, as a function of inverse wavenumber $1/k_{\phi_1}$ in the $\phi_1$ coordinate. The uncertainties on the power spectrum are derived by re-running the procedure on 1000 randomly-drawn profiles consistent with the profile uncertainties. The blue and green lines show the power spectra of the Gaia+Pan-STARRS and Gaia-only profiles (from Figure~\ref{fig:spikes}a and \ref{fig:spikes}b, respectively), while the yellow line is the power spectrum derived from the {\tt STREAMFINDER} sample. The three samples show similar, but not identical, behavior.

\begin{figure*}
\begin{center}
\includegraphics[angle=0, viewport= 15 5 855 425, clip, width=\hsize]{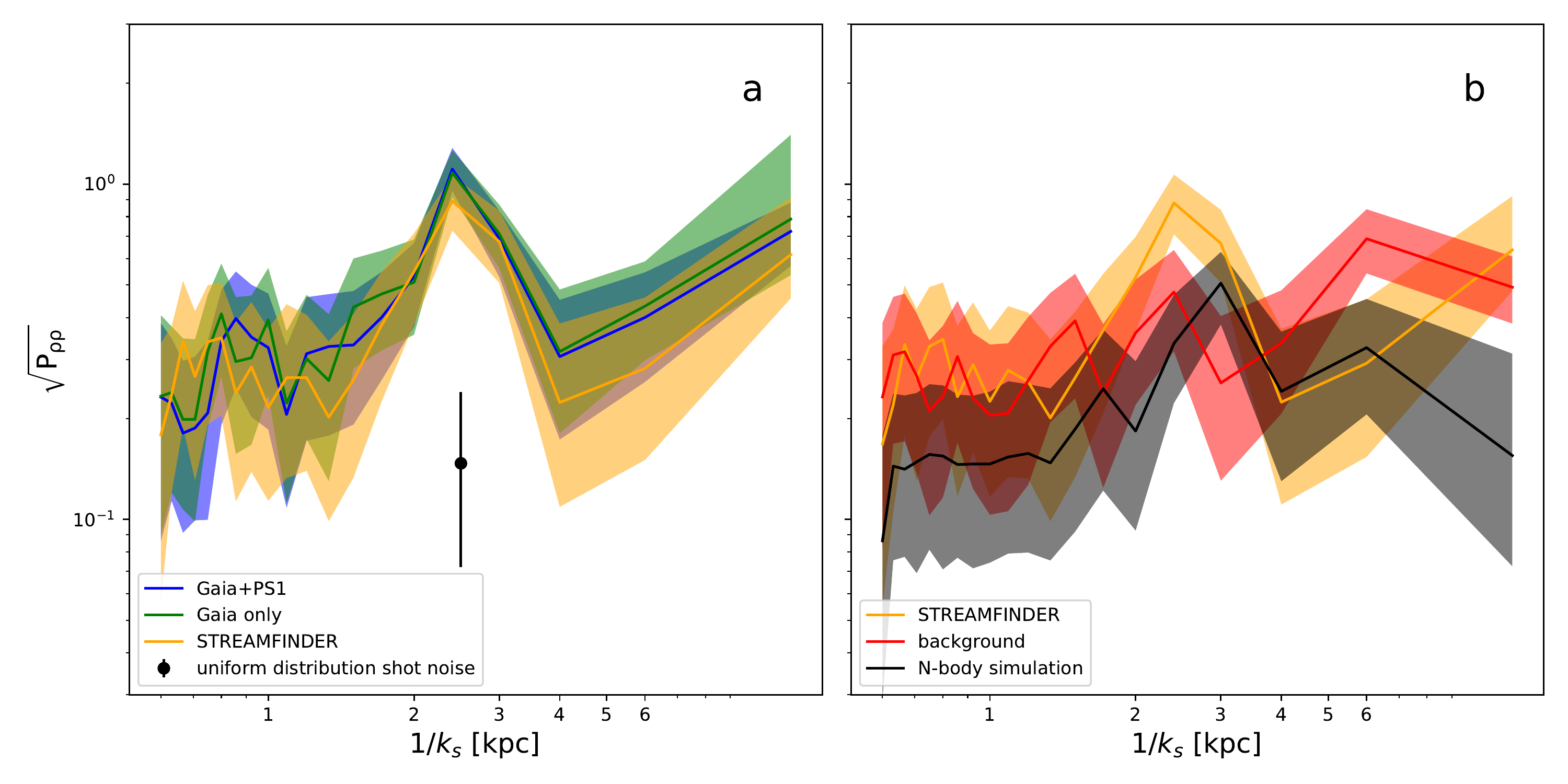}
\end{center}
\caption{As Figure~\ref{fig:powerspectrum}, but as a function of distance $s$ along the stream, rather than observed angle $\phi_1$. The fact that the peak at $2.64\pm0.18\kpc$ has much higher contrast here implies that projection effects smear out the periodic signal in Figure~\ref{fig:powerspectrum}. The black errorbar in (a) shows the $1\sigma$ uncertainty level due to shot noise in an artificial stream with 868 stars uniformly distributed in $s$.}
\label{fig:powerspectrum_s}
\end{figure*}

To serve as a comparison, we also calculate the power spectrum of the Galactic halo contamination in this region of sky (red line). For this, we chose to use the contamination profile previously shown in Figure~\ref{fig:spikes}a (blue histogram), derived from the sky region with $|\phi_2-S(\phi_1)|>2\deg$ which contains 861 stars in the range $\phi_1=[-90,10]$, an almost identical number to the Gaia+Pan-STARRS GD-1 sample (868 stars). The power spectrum of the smooth halo contaminants can be seen to be very similar to that of the {\tt STREAMFINDER} sample.

The measured power at different angular scales may come from clear structures, such as the spikes seen in Figure~\ref{fig:spikes}, and some signal may be due to interactions with invisible dark matter sub-halos. However, it is also possible that Gaia's scanning law (Figure~\ref{fig:Gaia_artifacts}) has imprinted a signal on the power spectra, and the fact that the background shows a similar density power spectrum to GD-1 lends weight to this concern. Indeed, the power spectra of the background and {\tt STREAMFINDER} samples fully overlap within $1\sigma$ over all spatial scales probed. Note in particular that the background sample extends over a very much wider range in the $\phi_2$ coordinate than the GD-1 stream samples, which was necessary in order to extract a similar number of stars. This means that the imprint of the scanning law will probably be diminished in the background sample since any artifacts will be averaged-out over the large $\phi_2$ range.

In Figure~\ref{fig:powerspectrum} we also show the power spectrum of the N-body model presented in Section~\ref{sec:Modelling_the_Density_Profile}. Despite the fact that the model has been integrated in a perfectly smooth Galactic potential, the power spectrum displays a considerable similarity to the observed profiles.

The slight bump in the power spectra at $\sim 15\deg$ in Figure~\ref{fig:powerspectrum}a is due to the strong peaks seen in Figure~\ref{fig:spikes}. Because of projection effects, the periodicity of the features is somewhat veiled when they are examined in the $\phi_1$ coordinate, but it becomes obvious after changing coordinates to a proper path length along the stream. Defining 
\begin{equation}
s(\phi_1)=\int_0^{\phi_1} \sqrt{D^2 + \Big({{d D}\over{d \phi^\prime_1}}\Big)^2} \, d \phi^\prime_1 
\end{equation}
to measure this path length, we re-calculate the density power spectra as a function of inverse wavenumber $1/k_s$ in this $s$ coordinate, and show the results in Figure~\ref{fig:powerspectrum_s}. The epicyclic over-densities produce a very clear signal at $1/k_s=2.64\pm0.18\kpc$ (calculated by fitting a Gaussian to the 1000 random realisations of the {\tt STREAMFINDER} sample, selecting data in the range $1/k_s=[1.3,4.5]\kpc$), and all three samples shown in Figure~\ref{fig:powerspectrum_s}a are consistent with each other. In Figure~\ref{fig:powerspectrum_s}b, we now see that the background sample does differ from the {\tt STREAMFINDER} sample, but mostly by the fact that it does not exhibit a pronounced peak at $1/k_s=2.64\pm0.18\kpc$. At small separations (where \citealt{2019arXiv191102662B} find that the perturbing influence of LCDM substructures is required), the background and {\tt STREAMFINDER} samples have identical behavior.

To interpret these power spectra it is useful to know the level of the shot noise. To estimate this, we made 1000 realisations of a uniform distribution in $s$ containing 868 stars, and calculated the corresponding power spectra. The resulting distributions are flat in $1/k_s$ and possess mean and $1\sigma$ uncertainties as displayed in Figure~\ref{fig:powerspectrum_s}a (black errorbar).

We note in passing that the larger epicyclic peak distance   in our N-body model (black line in Figure~\ref{fig:powerspectrum_s}b) suggests that the N-body model was $\sim 30$\% too massive at the time when the peaks were formed (estimated from Eq.~\ref{eq:Mv0rel}). This hint will be explored in future work.

\section{Conclusions}
\label{sec:Conclusions}

We have used the {\tt STREAMFINDER} algorithm to isolate a sample of stars of the GD-1 stellar stream over $\sim 100\deg$ of the Northern sky. Our radial-velocity follow-up of these {\tt STREAMFINDER} candidate members shows that there is very little contamination ($\sim 10$\%) if the sample is spatially restricted to being close ($<0.6\deg$) to the fitted path of GD-1. We take advantage of this sample to fit empirical relations to the sky position, radial velocity, proper motion and distance to the structure, which are then used to extract two other clearly-defined samples of GD-1 stars from the Gaia DR2 and Pan-STARRS catalogs. 

We find that the three different GD-1 samples we have constructed have a similar spatial distribution. In particular, very strong peaks are present along the stream, spaced by $2.64\pm0.18\kpc$, with a contrast exceeding 3:1.

While the density power spectrum may in principle contain information about the prevalence of perturbing massive bodies, for spatial separations up to $2\kpc$, we find a very similar behavior between the GD-1 samples we examined and a background profile extracted from the sky regions immediately adjacent to GD-1.

We also present a comparison between star-counts profiles derived from Pan-STARRS DR2 and a deeper survey with the MegaCam wide-field camera at the Canada-France Hawaii Telescope taken in much better seeing conditions. In a color-magnitude region where the GD-1 stream has highest contrast over the contaminants, we find substantial spatially-dependent differences in the corresponding star counts. We attribute these differences to variations in the observing conditions, leading to variations in completeness and variations in star/galaxy discrimination between the two imaging surveys. These errors are very hard to identify and correct for in large ground-based surveys without an external deeper dataset, and may strongly affect conclusions of the prevalence of density gaps in streams.

In contrast, space-based surveys may be more powerful for measuring density profiles, because they are unaffected by our variable weather. However, with a scanning instrument like Gaia, there may be spatially varying incompleteness due to the way in which the survey has been designed to cover the sky. In the particular region around GD-1 (Figure~\ref{fig:Gaia_artifacts}) there is evidently significant incompleteness on a range of spatial scales.

Nevertheless, the strong peaks detected here (Figures \ref{fig:spikes} and \ref{fig:map_density}) are clearly real, being also visible in the matched filter maps presented in the discovery paper \citep{2006ApJ...643L..17G}. Our modelling shows that these features are most probably due to epicyclic motion in the stream. This conclusion is strengthened by the finding of a strong periodic signal in the density power spectra when we correct for the projection effects (Figure~\ref{fig:powerspectrum_s}). The fact that these periodic density variations are still visible, and have not yet been washed out (as stars mix over time due to the dynamical evolution of the stream) implies that the progenitor of the system went through its final disruption stage only very recently.

In order to obtain reliable constraints of the effect of dark matter substructures from the density profiles of GD-1 it will be necessary to fully account for these internal dynamical properties of the stream, as well as the external perturbations from the baryonic components of the Milky Way. Finally, the (probably very complicated) instrumental sensitivity function will need to be corrected for.

\acknowledgments

We would like to thank the anonymous referee for several very insightful comments and suggestions.

RI, BF, NM and GM acknowledge funding from the Agence Nationale de la Recherche (ANR project ANR-18-CE31-0006, ANR-18-CE31-0017 and ANR-19-CE31-0017), from CNRS/INSU through the Programme National Galaxies et Cosmologie, and from the European Research Council (ERC) under the European Unions Horizon 2020 research and innovation programme (grant agreement No. 834148).

Based on observations obtained at the Canada-France-Hawaii Telescope (CFHT) which is operated by the National Research Council of Canada, the Institut National des Sciences de l´Univers of the Centre National de la Recherche Scientique of France, and the University of Hawaii.

Based on observations obtained with MegaPrime/MegaCam, a joint project of CFHT and CEA/DAPNIA, at the Canada-France-Hawaii Telescope (CFHT) which is operated by the National Research Council (NRC) of Canada, the Institut National des Science de l'Univers of the Centre National de la Recherche Scientifique (CNRS) of France, and the University of Hawaii.

Based on observations collected at the European Southern Observatory under ESO programme 103.B-0568(A) and 103.B-0568(B).

This work has made use of data from the European Space Agency (ESA) mission {\it Gaia} (\url{https://www.cosmos.esa.int/gaia}), processed by the {\it Gaia} Data Processing and Analysis Consortium (DPAC, \url{https://www.cosmos.esa.int/web/gaia/dpac/consortium}). Funding for the DPAC has been provided by national institutions, in particular the institutions participating in the {\it Gaia} Multilateral Agreement. 

Funding for SDSS-III has been provided by the Alfred P. Sloan Foundation, the Participating Institutions, the National Science Foundation, and the U.S. Department of Energy Office of Science. The SDSS-III web site is http://www.sdss3.org/.

SDSS-III is managed by the Astrophysical Research Consortium for the Participating Institutions of the SDSS-III Collaboration including the University of Arizona, the Brazilian Participation Group, Brookhaven National Laboratory, Carnegie Mellon University, University of Florida, the French Participation Group, the German Participation Group, Harvard University, the Instituto de Astrofisica de Canarias, the Michigan State/Notre Dame/JINA Participation Group, Johns Hopkins University, Lawrence Berkeley National Laboratory, Max Planck Institute for Astrophysics, Max Planck Institute for Extraterrestrial Physics, New Mexico State University, New York University, Ohio State University, Pennsylvania State University, University of Portsmouth, Princeton University, the Spanish Participation Group, University of Tokyo, University of Utah, Vanderbilt University, University of Virginia, University of Washington, and Yale University.

Guoshoujing Telescope (the Large Sky Area Multi-Object Fiber Spectroscopic Telescope LAMOST) is a National Major Scientific Project built by the Chinese Academy of Sciences. Funding for the project has been provided by the National Development and Reform Commission. LAMOST is operated and managed by the National Astronomical Observatories, Chinese Academy of Sciences.

\bibliography{GD1_Gaps}
\bibliographystyle{aasjournal}

\end{document}